\documentclass[11pt,a4paper]{article}

\usepackage{color}
\usepackage{graphicx}
\usepackage{slashed}
\usepackage{multicol}
\usepackage{multirow}
\usepackage{youngtab}
\usepackage{array}
\usepackage[backend=bibtex,sorting=none,firstinits=true,hyperref=true,doi=false,maxnames=5]{biblatex}

\bibliography{References}

\renewbibmacro{in:}{%
  }%

\usepackage{amsmath} 
\usepackage{amssymb} 
\usepackage{amsfonts} 
\usepackage{bbm} 

\usepackage{a4wide} 

\usepackage{lmodern} 
\usepackage[T1]{fontenc} 
\usepackage[utf8]{inputenc} 
\usepackage{fixltx2e} 
\usepackage{microtype} 

\usepackage{hyperref}

\newcommand{\LL}{{\cal L}}

\begin{document}

\begin{flushright}
\small
MAD-TH-13-06\\
MPP-2014-6
\date \\
\normalsize
\end{flushright}

\vskip 1cm

\begin{center}
{\Large \bf Building a St\"uckelberg Portal}

\vspace*{0.5in} {\large Wan-Zhe Feng$^{1,2}$, Gary Shiu$^{1,3}$, Pablo Soler$^{1,3}$, and Fang Ye$^{1,3}$}
\\[.3in]
{\em $^1$ Center for Fundamental Physics and Institute for Advanced Study, \\
Hong Kong University of Science and Technology, 
     Hong Kong} \\
     {\em $^2$ Max--Planck--Institut f\"ur Physik (Werner--Heisenberg--Institut),
80805 M\"unchen, Germany}\\
     {\em $^3$ Department of Physics,
     University of Wisconsin,
     Madison, WI 53706, USA} 
\\[0.3in]
\end{center}

\vskip 0.5cm

\begin{center}
{\bf
Abstract}
\end{center}
\noindent
We construct explicit string theory models realizing the recently proposed ``St\"uckelberg Portal'' scenario, a framework for building $Z'$ mediation models without the need to introduce unwanted exotic matter charged under the Standard Model. This scenario can be viewed purely field-theoretically, although it is particularly well motivated from string theory. By analyzing carefully the St\"uckelberg couplings between the Abelian gauge bosons and the RR axions, we construct the first global intersecting brane models which extend the Standard Model with a genuine hidden sector, to which it is  nonetheless connected via $U(1)$ mass mixings. Utilizing the explicit models we construct, we discuss some broad phenomenological properties and experimental implications of this scenario such as $Z-Z'$ mixings, dark matter stability and relic density, and supersymmetry mediation. With an appropriate confining hidden sector, our setup also provides a minimal realization of the hidden valley scenario. We further explore the possibility of obtaining small $Z'$ masses from a large ensemble of $U(1)$ bosons. Related to the St\"uckelberg portal are two mechanisms that connect the visible and the hidden sectors, namely mediation by non-perturbative operators and the hidden photon scenario, on which we briefly comment.

\vfill

\newpage

\tableofcontents

\newpage

\section{Introduction}

Besides supersymmetry, additional $U(1)$ gauge symmetries  (see e.g. \cite{Rizzo:2006nw, Langacker:2008yv} for reviews) and axion-like particles (ALPs) \cite{Jaeckel:2010ni,Essig:2013lka} are among the most common features in extensions of the Standard Model (SM).
An extended Abelian  gauge sector 
 often arises in top-down constructions,
either as a result of the breaking of a higher non-Abelian symmetry (as in Grand Unified Theories)
or simply because what completes the SM at high energies is likely to be rich enough to accommodate additional $U(1)$ factors.
Indeed, the large rank gauge group typically found in string constructions makes such Abelian extensions inevitable \cite{Ibanez:2012zz,Kakushadze:1997mc, Blumenhagen:2005mu,Blumenhagen:2006ci,Marchesano:2007de}.
Similarly, ALPs are abundant in string compactifications with the number of them  determined by the 
topology of the internal space. Their shift symmetries may be approximate (perturbative) symmetries of the low energy theory, in which case the ALPs are the associated pseudo-Nambu-Goldstone bosons; or they may be a part of $U(1)$ gauge symmetries, in which case the axions are absorbed by gauge bosons through St\"uckelberg couplings.

In this regard, the St\"uckelberg mechanism is particularly interesting as it naturally combines these two recurrent themes in beyond the SM physics. Our aim here is to suggest phenomenological scenarios involving St\"uckelberg $U(1)$'s, motivated by lessons learnt from concrete string constructions.
At a formal level, the couplings between Abelian gauge fields and their associated ALPs 
are essential for cancelling the apparent anomalies in the low energy spectrum. Nevertheless, the $U(1)$'s
that pick up a St\"uckelberg mass are not limited to those that are anomalous,
and the St\"uckelberg coupling has wider applicability in particle physics which we shall explore in the present work. 
 
From a phenomenological viewpoint, these St\"uckelberg $U(1)$'s  are also special in that they can provide an intriguing portal into dark sectors. 
General symmetry principles restrict the allowed interactions of the SM. One might argue that the Higgs boson $H$ is unique because, with the exception of the Higgs mass term $\mu^2 H^{\dagger} H$, the couplings in the SM are all strictly renormalizable. Given such a term, it is easy to construct a renormalizable operator between $H$ and a hidden sector scalar field $\phi_h$, namely $\phi_h^{\dagger}\phi_h H^{\dagger}H$, which can serve as an efficient portal \cite{Patt:2006fw}. Thus, the Higgs boson may be the only SM field that has renormalizable couplings with hidden sector fields.

This Higgs portal, as well as other proposals such as the vector, axion and neutrino portals, have been the object of recent intense investigation~\cite{Essig:2013lka}.
Here, we point out that the generic appearance of  extra $U(1)$ gauge fields $A_v$ and $A_h$ in both the visible and the hidden sectors, may provide yet another efficient 
gateway into dark sectors. In the presence of such bosons,
in addition to the Higgs portal, there are two other renormalizable couplings allowed by the symmetries that connect the visible and the hidden sectors, namely $m^2_{vh} A_v A_h$ (mass mixing) and $g_{vh}F_v F_h$ (kinetic mixing).
Any $U(1)$ under which the SM is charged (e.g. baryon or lepton number, Peccei-Quinn symmetries, etc) can mix with the hidden $U(1)$'s through these operators. 
 Kinetic mixing between the visible and hidden sectors is typically small as such effect is loop generated. As we shall argue, there exist well-motivated  {\it tree-level} mass mixing effects between $U(1)_v$ and $U(1)_h$.
 After diagonalization, the physical $Z'$ gauge bosons (in the sense that they have diagonal kinetic and mass terms) 
 will be linear combinations of $A_v$ and $A_h$ and can couple with significant strength to both visible and hidden matter fields simultaneously. 
The mass mixing between visible and hidden sector $U(1)$'s may therefore be the dominant channel for communication between the separated sectors.

Furthermore, as the $U(1)$ anomalies can be canceled by the Green-Schwarz (GS) mechanism, no exotic matter fields are needed to add in  for consistency.
Thus, the ``St\"uckelberg portal'' proposed here is perhaps one of the minimal extensions of the SM to 
SM$\oplus$Dark Sector.
As we will show, this minimal scenario is also naturally realized in string theory.
As already mentioned, among the generic features of string (and in particular, D-brane) constructions of particle physics models is the unavoidable presence of multiple St\"uckelberg 
$U(1)$'s under which the SM particles are charged.
Mass mixing of these visible $U(1)$'s with those of the hidden sector may be implemented through simple topological conditions, and provides a concrete realization of  our St\"uckelberg portal scenario.

Our scenario has certain similarities with some previous proposals in the literature  but there are also important differences. Mass mixing of hidden sector $U(1)$'s  with the SM hypercharge has been considered before \cite{Dermisek:2007qi,Verlinde:2007qk, Kors:2004dx}. However, after such a mixing, charged matter in the hidden sector (if present) would generically acquire a sizeable fractional electric charge~\cite{Shiu:2013wxa} which is extremely hard to reconcile with observational exclusion bounds~\cite{Langacker:2011db}.
Even if such matter is not present in the hidden sector, the mixing of extra $U(1)$'s with hypercharge would be highly constrained by electro-weak precision measurements.

Our proposal is perhaps closest in spirit to  $Z'$ mediation \cite{Langacker:2007ac} and to 
hidden valley scenarios \cite{Strassler:2006im}, in which a $Z'$ boson is often suggested as a possible mediator
between the visible and hidden sectors. 
When one attempts to construct explicit realizations of these setups, a strong mixing between the hidden and visible sectors (which allows the $Z'$ bosons to couple with sizeable strength to both sectors) often comes with unnecessary exotic matter fields. One may view our scenario as a way to generate the desired mixings in a simple manner without introducing the unwanted exotic matter.

As an illustration of our scenario, we present the first explicit global D-brane constructions which extend the Standard Model sector with a genuine hidden sector (i.e., with no exotic matter charged under the hidden sector that simultaneously carry SM charges) but nonetheless admit strong mixings between them. We expect the ingredients we developed in our explicit string constructions with the aforementioned features to have useful applications to other hidden sector physics as well.

This paper is organized as follows.
In Section~\ref{StringConstructions} we review basic features of SM-like constructions with intersecting branes, paying special attention to the St\"uckelberg couplings between Abelian gauge bosons and Ramond-Ramond (RR) axions. Using the ingredients presented there, we describe in Section~\ref{sec:portal} our generic framework and the mechanism by which the St\"uckelberg portal (i.e. $Z'$ mediation into a hidden sector) can be implemented in type IIA compactifications. We also present there an explicit toroidal construction that shows most of the generic features we are interested in. In section~\ref{sec:pheno} we study some of the phenomenological properties and experimental implications of the stringy St\"uckelberg portal scenario, and perform explicit computations for the mentioned toroidal model. In Section~\ref{sec:random} we discuss the relation between the $Z'$ boson masses and the string scale. We briefly review known mechanisms to obtain $Z'$ masses in a phenomenologically interesting range, and study the possibility of obtaining small masses from a large ensemble of $U(1)$ bosons. In Section~\ref{relatedportals} we comment on two mechanisms, different but related to the St\"uckelberg portal, that connect visible and hidden sectors, namely non-perturbative effects and kinetic mixings among axions. In Section~\ref{conclusions} we present our conclusions.

The main ideas of this paper have been outlined in the short note~\cite{short}, which takes a field theoretical approach. In this paper we present a more detailed analysis of the St\"uckelberg portal scenario, and focus specially on its natural implementation in string theory setups.

\section{$U(1)$'s in String Constructions}\label{StringConstructions}

In this section we review the construction of particle physics models from type II string theories, paying special attention to the properties and roles played by $U(1)$ gauge symmetries. For an overview of string theoretical constructions of particle physics, see~\cite{Ibanez:2012zz,Blumenhagen:2005mu,Blumenhagen:2006ci,Marchesano:2007de} and references therein. We will focus here for simplicity on type IIA models of intersecting D6-branes, although most of our discussion can be applied to other type II setups such as branes at singularities or magnetized branes, as expected from string dualities, with an appropriate reinterpretation of the ingredients involved.

\subsection{Models of intersecting D6-branes}\label{setup}
Some of the simplest and most intuitive, yet most successful, constructions of chiral effective gauge theories in string theory are obtained by considering stacks of D6-branes intersecting at angles in type IIA orientifold compactifications. We briefly review here these constructions and introduce some of the notation we will be using throughout this work.

Given a four dimensional type IIA compactification, gauge theories arise from stacks of D6-branes that wrap three-cycles of the internal manifold $\bf{X_6}$, usually taken to be Calabi-Yau (C.Y.), and span the four non-compact Minkowskian directions.

In compact models, tadpoles introduced by the D6-branes are cancelled by Orientifold 6-planes (O6-planes) which carry $-4$ units of D6-brane charge. Correspondingly one must include image D6-branes so that the system is invariant under the orientifold projection. Tadpole cancellation requires the homologies of the three-cycles wrapped by the branes and the O6-plane to satisfy
\begin{equation}\label{tadpoles}
\sum_aN_a[\Pi_a]+\sum_aN_a[\Pi_a]'-4\,[\Pi_{O6}]=0\,,
\end{equation}
where $N_a$ is the number of coincident D6-branes in the stack $a$ which wraps a three-cycle in the integer homology class $[\Pi_a]\in H_3({\bf{X_6}},\mathbb{Z})$, whose orientifold image we denote $[\Pi_a]'$.
Notice that, since both branes and image-branes contribute to the tadpoles, only homology classes that are even under the orientifold projection enter this relation. 

The 4d effective theory of open strings living on a stack of $N$ coincident D6-branes is given by a gauge theory with gauge group $U(N)$. Crucially for our purposes, such a group is (locally) $U(N)\cong SU(N)\times U(1)$, and contains an abelian $U(1)$ factor with which we will be mostly concerned.\footnote{Globally $U(N)\cong [SU(N)\times U(1)]/\mathbb{Z}_N$, where $\mathbb{Z}_N\subset U(1)$ is identified with the center of $SU(N)$. Such a subtlety will not play a role in our discussion.} 
If a stack of $N$ branes coincides with its own image (i.e. $\Pi_a=\pm\Pi_a'$), the associated gauge group is not $U(N)$ but rather $SO(N)$ or $USp(N)$. Since such groups do not contain abelian factors, which are our main object of study in this work, they will not play a role in our discussion.

In the simplest cases, we can introduce a basis of three-cycles $\{[\alpha^i],[\beta_i]\}_{i=0,\ldots,h_{2,1}}$ of ${\bf{X_6}}$, with $[\alpha^i]$ even and $[\beta_i]$ odd under the the orientifold projection, whose topological intersection numbers read
\begin{equation}\label{basis}
[\alpha^i]\cdot[\beta_j]=-[\beta_j]\cdot[\alpha^i]=\delta_j^i\,.
\end{equation}
One can then express the cycles wrapped by a given stack of branes in terms of this basis as
\begin{equation}
[\Pi_a]=s_{ai}[\alpha^i]+r_a^{\,j}[\beta_j]\qquad\qquad
[\Pi_a]'=s_{ai}[\alpha^i]-r_a^{\,j}[\beta_j]\,,
\end{equation}
where the coefficients $s_{ai}$ and $r_a^{\,j}$ are integer wrapping numbers. 

A slightly more general situation, which we will use in section \ref{Madrid} and is sometimes referred to as `tilted orientifold' (in analogy to tilted tori), occurs when some of the cycles of the basis do not have a definite parity under the orientifold, but rather transform as, e.g.
\begin{equation}
[\tilde\alpha^i]\to[\tilde\alpha^i]-[\tilde\beta_j]\,,~~~[\tilde\beta_j]\to-[\tilde\beta_j]\,,~~~~~~{\text{for some  }}i,j\,.
\end{equation}
In that case, one can still define new cycles $[\tilde\alpha^i]\equiv [\alpha^i]-1/2[\beta_j]$ and $[\tilde\beta_j]\equiv[\beta_j]$, that are even and odd under the orientifold, respectively, and that still satisfy~\eqref{basis}. The only subtlety when expressing general cycles in terms of the new basis $\{[\tilde\alpha^i],[\tilde\beta_i]\}$, is that some of the coefficients in the expansion might be half-integers:
\begin{equation}
[\Pi_a]=s_{ai}[\alpha^i]+r_a^{\,j}[\beta_j]=s_{ai}[\tilde\alpha^i]+\underbrace{(r_a^j +\tfrac{1}{2}\,s_{ai})}_{\in \,\mathbb{Z}/2}[\tilde\beta_j]
\end{equation}
and one has to be careful in keeping track of possible factors of 2 in computations. 

At the intersection of two brane stacks $a$ and $b$ 
there are massless chiral fermions that transform under the bifundamental representation $(\,\tiny\yng(1)_a\,,\overline{\tiny{\yng(1)}}_{\,b}\,)_{(+1,-1)}$ of the gauge group $SU(N_a)\times SU(N_b) \times U(1)_a\times U(1)_b$, and come from open strings that stretch from stack $a$ to stack $b$. The number of replicas of such chiral fermions is given by the topological intersection number $[\Pi_a]\cdot[\Pi_b]$ which reads
\begin{equation}\label{intersections}
[\Pi_a]\cdot[\Pi_b]=s_{ai}\,r_b^i- r_a^{\,i}\,s_{bi}\,.
\end{equation}
At the intersection of a stack of branes with its own image, matter charged under more general representations of $U(N)$, such as the symmetric or antisymmetric, may also arise.

\subsection{St\"uckelberg $U(1)$ masses}\label{sec:stuck}
Besides these fermions, the open string $U(1)$ gauge bosons also couple to closed string RR axions $\phi^i$ that pair up with the geometric complex structure moduli $u^i$ of the compactification. These fields arise from the reduction of the holomorphic three form $\Omega_3$ of the C.Y. and the RR three-form $C_3$ along internal three-cycles as
\begin{equation}\label{csmoduli}
U^i\equiv u^i+i\phi^i=\int_{[\alpha^i]} \text{Re}(C\Omega_3)+i\int_{[\alpha^i]} C_3\qquad\qquad i=0,\ldots, h_{2,1}
\end{equation}
where $C$ is a normalization factor proportional to $C\propto e^{-\phi}=1/g_s$. We have included in this definition the complex dilaton as $U^0=S$. The RR axions $\phi^i$ are periodic, and in the appropriate normalization one can identify $\phi^i\sim \phi^i+1$.

It turns out that these axions undergo non-trivial shift transformation under the $U(1)_a$ gauge symmetries introduced above:
\begin{equation}\label{axionshift}
A^a\to A^a+d\Lambda^a\qquad\qquad \phi^i\to \phi^i+N_a \,r_a^{\,i}\Lambda^a\,.
\end{equation}
Accordingly, the kinetic terms of the complex structure moduli $U^i$ must include derivatives that are covariant under these shifts
\begin{eqnarray}\label{St\"uckelberg}
\LL_{\text{kin}}&=&-\frac{1}{2}G_{ij}(D_\mu U^i)^\dagger( D^\mu U^j) \nonumber\\
&=&-\frac{1}{2}G_{ij}\left(\partial_\mu u^i \partial^\mu u^j\right) -\frac{1}{2}G_{ij}\left(\partial_\mu \phi^i-N_a\,r_a^{\,i}\, A^{a}_{\mu}\right)\left(\partial^\mu \phi^{\,j}-N_b\,r_b^{\,j} A^{b\mu}\right)\,,
\end{eqnarray}
where $G_{ij}$ is the (positive definite) metric on the complex structure moduli space. We see that this Lagrangian encodes a St\"uckelberg mechanism by which the $U(1)$ gauge bosons gain masses through the absorption of the RR-axions $\phi^i$.\footnote{The couplings between the axions and the gauge bosons can be seen to arise from the D6-brane Chern-Simons action $S_{\text{CS}}\sim N_a\int_{D6_a}C_5\wedge {\text{tr}}F_a\to N_a\,r_a^{\,i}\int_{4\text{d}}B_i\wedge \text{tr}F_a$, where $B_i\equiv\int_{[\beta_i]}C_5=\ast_{4\text{d}}\phi^i$ is the Hodge dual of the complex structure axions $\phi^i$.} 

Let us note that the structure of the gauge boson-axion couplings of~\eqref{St\"uckelberg} plays a crucial role in the GS mechanism that cancels the triangle anomalies of anomalous $U(1)$'s. The tadpole cancellation condition~\eqref{tadpoles} guarantees that the full system is indeed anomaly-free. However, the St\"uckelberg mechanism is not restricted to the anomalous $U(1)$'s, and even gauge bosons of non-anomalous factors may acquire a mass. 

The mass matrix of $U(1)$ bosons can be read straightforwardly from~\eqref{St\"uckelberg}:
\begin{equation}\label{masses}
M_{ab}^2=N_a\,r_a^{\,i}\,G_{ij}\,r_b^{\,j} N_b\equiv \left(K^{\,T}\cdot G\cdot K\right)_{ab}
\end{equation}
where we have defined the rectangular matrix with integer entries $(K)^i_a=N_a r_a^{\,i}$, which basically encodes the linear combinations of odd cycles $[\beta_i]$ wrapped by the branes $[\Pi_a]$. Alternatively, it can be interpreted as the matrix of $U(1)_a$ charges of the composite fields $e^{\,U^i}$ that transform linearly under the $U(1)$ symmetries.\footnote{Incidentally, vacuum expectation values (vev's) of these composite fields $\langle e^{\,U^i}\rangle$ are responsible for the breaking of the massive $U(1)$'s (c.f. section \ref{sec:non-perturbative}), so they indeed behave similarly to Higgs fields.\label{footnote}} We will often refer to the entries of $K$ as `axion charges'.

It is clear from eq.~\eqref{St\"uckelberg} that the gauge bosons which remain massless are those that do not couple to any RR axions. This will be the case for linear combinations $A_{\vec{v}}=v^a A_a$ such that 
\begin{equation}\label{massless}
N_a\,r_a^{\,i}\,v^a=0\qquad\text{equiv.}\qquad K\cdot\vec{v}=0\,.
\end{equation}
$\vec{v}$ is a vector in the space of $U(1)$'s which, given the fact that $K$ is integral, can always be chosen to have integer entries. Hence, the number of gauge bosons that acquire a mass is equal to the rank of $K$, or equivalently, to the number of fields $e^{\,U^i}$ with linearly independent charge vectors. Notice also that the masslessness condition~\eqref{massless} has a nice interpretation in terms of the odd-homology $H^-_3({\bf{X_6}},{\mathbb{Z}})$: massless gauge fields correspond to linear combinations of branes that wrap trivial cycles in this odd homology, i.e. $v^a\,N_a([\Pi_a]-[\Pi_a]')=0$.

\subsection{The Higgs mechanism}\label{sec:higgs}
We can also consider giving a mass to certain $U(1)$'s through the usual Higgs mechanism, i.e. by giving a vacuum expectation value to open string scalar fields $H^j=h^je^{i\phi^j}$ charged under them. This process has an interpretation in terms of the recombination of the corresponding branes as explained in \cite{Cremades:2002cs}. The effective Lagrangian takes the form
\begin{eqnarray}
\LL_{\text{Higgs}}&=&-\frac{1}{2}\,\delta_{ij}(D_\mu H^i)^\dagger (D^\mu H^j)\nonumber\\
&=&-\,\frac{1}{2}\delta_{ij}\,\partial_{\mu}h^i\partial^{\mu} h^j-\frac{1}{2}\, \delta_{ij}(h^i)^2\left(\partial_\mu \phi^i-q_a^{\,i} A^a_\mu\right)\left(\partial^\mu \phi^j-q_b^{\,j} A^{b\mu}\right)\,,
\end{eqnarray}
which is  the same as~\eqref{St\"uckelberg} under the identifications $K_a^{\,i}\sim q_a^{\,i}$ and $G_{ij}\sim (h^i)^2\delta_{ij}$ (for simplicity we have considered the canonical metric in the space of Higgs fields). The $U(1)$ charges of the Higgs fields are encoded in the matrix $(q)_a^{\,i}$ which also has integer entries.

If the Higgs fields have also non-abelian charges, we can still write the same Lagrangian by including among the $U(1)$ gauge fields $A^a$ the components of the non-abelian groups that mix with the abelian factors (the Cartan components), e.g. the third component of $SU(2)_L$ in the usual EWSB of the SM. We will still refer to these components as $U(1)$ gauge bosons despite coming from non-abelian groups.

Summarizing, we can encode the $U(1)$ mass matrix coming from both the Higgs and the St\"uckelberg mechanism as in~\eqref{masses}:
\begin{equation}
M^2=K^{\,T}\cdot G\cdot K
\end{equation}
where $K$ is an integer matrix that encodes the $U(1)$ charges of the RR axions and the Higgs fields, while $G$ is a positive definite matrix that encodes the complex structure moduli space metric and the vev's of the Higgs fields. The eigenvalues of $M^2$ are non-negative, and zero eigenvectors $\vec{v}$ satisfy the condition $K\cdot \vec{v}=0$.

\subsection{A SM-like construction}\label{Madrid}

Let us wrap up this section by reviewing one of the well known semi-realistic construction of intersecting D6-branes~\cite{Ibanez:2001nd}, which will serve as a starting point for our explicit models of section~\ref{model}. Most of the discussion can be presented at a topological level, what is sometimes referred to as {\it proto}-models, postponing the actual embedding into a particular geometry to a later stage. 

In order to reproduce the SM gauge group and chiral matter, one introduces four stacks of branes (and their orientifold images) that lead to a gauge group $U(3)_a\times U(2)_b\times U(1)_c\times U(1)_d$ with the following intersection numbers:
\begin{eqnarray}\label{intersm}
&[\Pi_a]\cdot[\Pi_b]=1\,,~[\Pi_a]\cdot[\Pi_b]'=2\,;~~~\,~~~~~
&[\Pi_a]\cdot[\Pi_c]=-3\,,~[\Pi_a]\cdot[\Pi_c]'=-3\,; \nonumber\\
&[\Pi_b]\cdot[\Pi_d]=0\,,~[\Pi_b]\cdot[\Pi_d]'=-3\,;~~~~~~
&[\Pi_c]\cdot[\Pi_d]=-3\,,~[\Pi_c]\cdot[\Pi_d]'=3\,;
\end{eqnarray}
other intersections being zero. These numbers yield a chiral spectrum described in table~\ref{IMR}, where hypercharge is given by the linear combination
\begin{equation}\label{hyper}
Q_Y=\frac{1}{6}(Q_a-3Q_c+3Q_d)\,.
\end{equation}

\begin{table}[htb] \footnotesize
\renewcommand{\arraystretch}{1.25}
\begin{center}
\begin{tabular}{|c|c|c|c|c|c|c|c|}
\hline Intersection &
 Matter fields  & $SU(3)\times SU(2)$  &  $Q_a$  & $Q_b $ & $Q_c $ & $Q_d$  & $Q_Y$ \\
\hline\hline (ab) & $Q_L$ &  $(3,2)$ & 1  & -1 & 0 & 0 & 1/6 \\
\hline (ab') & $q_L$   &  $2( 3,2)$ &  1  & 1  & 0  & 0  & 1/6 \\
\hline (ac) & $U_R$   &  $3( {\bar 3},1)$ &  -1  & 0  & 1  & 0 & -2/3 \\
\hline (ac') & $D_R$   &  $3( {\bar 3},1)$ &  -1  & 0  & -1  & 0 & 1/3 \\
\hline (bd') & $ L$    &  $3(1,2)$ &  0   & -1   & 0  & -1 & -1/2  \\
\hline (cd) & $E_R$   &  $3(1,1)$ &  0  & 0  & -1  & 1  & 1   \\
\hline (cd') & $N_R$   &  $3(1,1)$ &  0  & 0  & 1  & 1  & 0 \\
\hline \end{tabular}
\end{center} 
\caption{\small Standard Model spectrum and $U(1)$ charges corresponding to the intersection numbers in (\ref{intersm}), reproducing the models in~\cite{Ibanez:2001nd}.}
\label{IMR} 
\end{table}

Scalar fields with the quantum numbers of the Higgs boson arise from open strings that stretch between the parallel stacks $b$ and $c$, and/or their orientifold images, which may be light if these branes are close enough to each other.

The intersection numbers~\eqref{intersm} (in particular $[\Pi_a]\cdot[\Pi_b]$ and $[\Pi_a]\cdot[\Pi_b]'$) require considering the `tilted orientifolds'  described in section~\ref{setup}. We will always work in the basis $\{[\tilde\alpha^i],[\tilde\beta_j]\}$ whose elements have definite parities under the orientifold, and from now on, will drop their tildes. As mentioned before, in this basis there might arise half-integer wrapping numbers. 

Given an appropriately tilted orientifold with $h_{2,1}\geq 3$, and assuming the O6-planes wrap around $[\alpha_0]$, the following wrapping numbers reproduce the intersections~\eqref{intersm}, and hence, the SM spectrum of Table~\ref{IMR}:\footnote{This is just a subclass of the models constructed in~\cite{Ibanez:2001nd}, obtained by setting $\epsilon=\rho=\beta^1=\beta^2=1$.}
\begin{align}
[\Pi _{a}]&=n_{a}[\alpha ^{0}]+\frac{1}{2}[\alpha ^{1}]+[\beta _{2}]+\frac{n_{a}}{2}[\beta _{3}],\nonumber\\
[\Pi _{b}]&=n_{b}[\alpha ^{0}]-\frac{3}{2}[\alpha ^{2}]-[\beta _{1}]+\frac{3n_{b}}{2}[\beta _{3}],\nonumber\\
[\Pi _{c}]&=3[\alpha ^{2}]+n_{c}[\beta _{3}],\nonumber\\
[\Pi _{d}]&=n_{d}[\alpha ^{0}]-\frac{3}{2}[\alpha ^{1}]-[\beta _{2}]+\frac{3n_{d}}{2}[\beta _{3}],\label{wrappings}
\end{align}
where $n_a,\,n_b,\,n_c,\,n_d\in\mathbb{Z}$. Hypercharge~\eqref{hyper} will be massless as long as 
\begin{equation}\label{masslesshyper}
n_a-2n_c+3n_d=0\,.
\end{equation}
Finally, if the number of O6-planes along $[\alpha^0]$ is $N_{O6}$, tadpole cancellation requires
\begin{equation}
3n_a+2n_b+n_d=2N_{O6}\,.
\end{equation}

Already from~\eqref{wrappings} one can read the matrix $K$ involved in the matrix of $U(1)$ masses~\eqref{masses}:
\begin{equation}\label{K-IMR}
K=\left(
\begin{array}{cccc}
0&0&0&0\\
0&-2&0&0\\
3&0&0&-1\\
\frac{3n_a}{2}&3n_b&n_c&\frac{3n_d}{2}
\end{array}\right)
\end{equation}
In order to reconstruct the full mass matrix, however, one needs the explicit form of the complex structure moduli space metric $G$, which depends on the particular geometry of the internal space in which this {\it proto}-model is implemented. It can be already seen from~\eqref{K-IMR}, however, that for generic $G$ (in particular for a diagonal metric), the mass matrix will be highly non-diagonal, and physical $Z'$ eigenvectors will be linear combinations of all the $U(1)$'s of the system. We will discuss this diagonalization in detail in the following sections.

\begin{figure}[!htp]
\centering
\includegraphics[width=440pt]{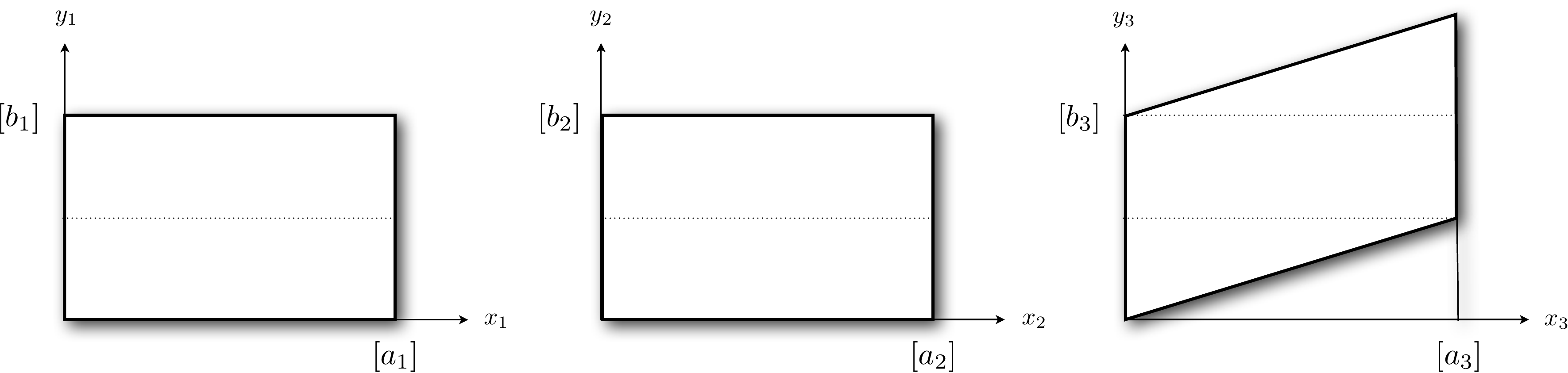}
\caption{\footnotesize Factorizeble six-torus $T^2_1\times T_2^2 \times T_3^2$ with $T_3^2$ tilted. Notice that $[a_3]$ does not represent a closed one-cycle on $T_3^2$, only linear combinations such as $2[a_3]$ or $[a_3]+\frac{1}{2}[b_3]$ do. This leads to the half-integer wrapping numbers in~\eqref{wrappings}. The antiholomorphic involution $(x_i,y_i)\to(x_i,-y_i)$ introduces four O6-planes along $2[\alpha^0]=[a_1]\times[a_2]\times2[a_3]$.}\label{tilted}
\end{figure}

This {\it proto}-model can be explicitly implemented in terms of factorizable cycles of a six torus $T^2_1\times T_2^2 \times T_3^2$, where the third one is tilted, see figure~\ref{tilted}. Given coordinates $(x_i, y_i)$ for each torus $(i=1,2,3)$, let the orientifold act as $(x_i, y_i)\to(x_i, -y_i)$, and let $[a_{i}]$ and $[b_{i}]$ be the even and odd $1$-cycles on $(T^{2})^{i}$. The basis $\{[\alpha^i,\beta_j]\}$ can then be expressed as
\begin{align}
[\alpha ^{0}]=[a_{1}][a_{2}][a_{3}],\,\,\,\,[\beta _{0}]=[b_{1}][b_{2}][b_{3}],\notag \\
[\alpha ^{1}]=[a_{1}][b_{2}][b_{3}],\,\,\,\,[\beta _{1}]=[b_{1}][a_{2}][a_{3}],\notag \\
[\alpha ^{2}]=[b_{1}][a_{2}][b_{3}],\,\,\,\,[\beta _{2}]=[a_{1}][b_{2}][a_{3}],\notag \\
[\alpha ^{3}]=[b_{1}][b_{2}][a_{3}],\,\,\,\,[\beta _{3}]=[a_{1}][a_{2}][b_{3}],
\end{align}

In this case, there are eight O6-planes along $[\alpha^0]$ (equivalently four O6-planes along $2[\alpha^0]$), so the tadpole cancellation condition reads
\begin{equation}
3n_a+2n_b+n_d=16\,.
\end{equation}

These models have been extensively studied in the literature. For the toroidal implementation, the complex structure moduli space metric $G$ is diagonal (at tree level), and the mass matrix of $U(1)$ can be fully determined. In particular, a thorough study of the  $U(1)$ gauge bosons and their masses was carried out in~\cite{Ghilencea:2002da}. In the following, we will generalize this analysis to include a hidden gauge sector whose $U(1)$'s mix with those from the visible one.

\section{$U(1)$ mass mixing and $Z'$ mediation}\label{sec:portal}
\begin{figure}[!htp]
\centering
\includegraphics[width=200pt]{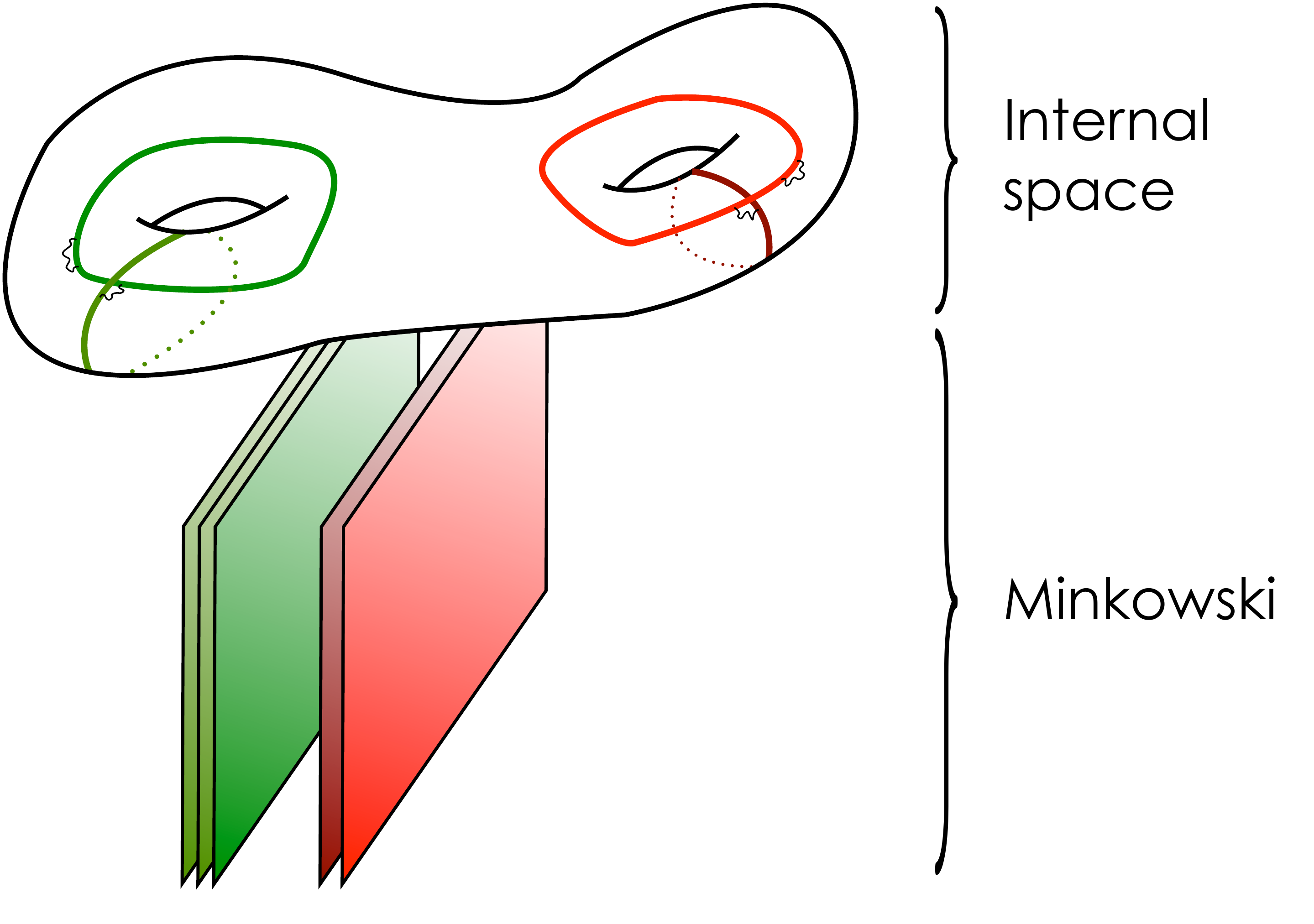}
\caption{\footnotesize Schematic representation of a hidden sector scenario~\eqref{stuckportal} with intersecting D-branes. The green and red stacks do not intersect each other and host different gauge and matter sectors.}\label{drawing}
\end{figure}

Our goal in this work is to study the role played by $U(1)$ gauge bosons and their mixings as portals into hidden sectors. We consider a generic setup, schematically depicted in Figure~\ref{drawing}, in which the visible sector consists of a SM-like construction, such as the one just described, and a hidden sector, whose branes do not intersect with the visible ones. 
The scenario we consider can be schematically written as
\begin{eqnarray}\label{stuckportal}
SU(3)_{c}\times SU(2)_{L}\times U(1)_{\text{v}}^n~~\times~~U(1)_{\text{h}}^m\times \tilde{G}_{\text{h}} \,,\\[-10pt]
\underbrace{ \hphantom{SU(3)_{c}\times SU(2)_{L}\times U(1)_{\text{v}}^n}}_{\Psi_{\text{v}}} \hphantom{~~\times~~}\underbrace{ \hphantom{U(1)_{\text{h}}^m\times\tilde{G}_{\text{h}}}}_{\Psi_{\text{h}}}\hphantom{.}\nonumber
\end{eqnarray}
where $\tilde{G}_{\text h}$ represent the semi-simple part of the hidden gauge group.

We can arrange the $U(1)$ gauge bosons in a vector $\vec{A}=(A_1^{\text{(v)}}\ldots A_n^{\text{(v)}}\,;\,A_{n+1}^{\text{(h)}}\ldots A_{n+m}^{\text{(h)}})$. The part of the effective Lagrangian that involves them can be written as
\begin{equation}\label{lag}
{\cal{L}}=-\frac{1}{4}\vec{F}^{\,T}\cdot f\cdot\vec{F}-\frac{1}{2} \vec{A}^{\,T}\cdot M^2\cdot\vec{A}+\sum_r\vec{q}_{r}^{\,T}\cdot\vec{A}\,J^{\,r}.
\end{equation}
At tree-level, $f=\text{diag}(g_1^{-2}\,\ldots\,g_{n+m}^{-2})$ is diagonal and encodes the $U(1)$ coupling constants, while non-diagonal terms can arise at loop-level by integrating out massive states charged under different $U(1)$'s~\cite{Holdom:1985ag}.

As explained in section~\ref{sec:stuck}, the mass matrix reads $M^2=K^{\,T}\cdot G\cdot K$, where $G$ is a positive definite metric, and $K$ is a matrix of integer entries (perhaps half-integers if the orientifold is tilted). Currents $J^{\,r}$ of matter fields couple to the $U(1)$ bosons with integer charges $\vec{q}_r$, encoded in the Chan-Paton indices of the open strings from which they arise.\footnote{The fact that one can always find a normalization of gauge fields in which $K$ and all the charges of the system are integral is not a feature specific to string theories, it is rather a simple consequence of the compactness of the gauge group. Theories with non-compact gauge groups, or equivalent with non-quantized charges, are incompatible with general {\it folk} theorems of quantum gravity~\cite{Banks:2010zn}.}

Mixing of $U(1)$'s from separated sectors have remarkably different qualitative features depending on whether it is induced by kinetic $(f)$ or mass $(M^2)$ terms, and on whether the gauge bosons involved are massive or massless. 

Kinetic mixing of massless hypercharge $U(1)_{\text{Y}}\subset U(1)^n_{\text v}$ with a hidden $U(1)_{\text h}$ has been thoroughly studied in the context of string theory~\cite{Lust:2003ky,Abel:2003ue,Berg:2004ek,Abel:2006qt,Abel:2008ai,Goodsell:2009xc,Gmeiner:2009fb,Cicoli:2011yh,Honecker:2011sm}. If the hidden gauge boson is massless, the mixing leads to the remarkable appearance of non-quantized small hypercharges in the hidden sector~\cite{Holdom:1985ag,Holdom:1986eq} (equivalently electric mini-charges after EWSB). On the other hand, if the hidden boson has a mass, kinetic mixing with hypercharge will induce a small coupling between the SM particles and the hidden $U(1)_{\text h}$. This coupling leads to very interesting phenomenological consequences for cosmology, astrophysics and particle physics, in setups in which the hidden photon is light enough (c.f. section~\ref{hiddenphotons}).

Off-diagonal kinetic terms are loop-suppressed effects and are expected to be quite small. We will rather focus in this work on mass mixing terms, which can appear already at tree level. Unless stated explicitly, we will neglect subleading kinetic mixing effects and assume that $f$ is diagonal.

\subsection{Mass mixing of $U(1)$ bosons}\label{sec:theory}
Mass mixing of hypercharge with hidden $U(1)$'s has been considered in~\cite{Kors:2004dx,Cheung:2007ut,Feldman:2007wj,Dermisek:2007qi,Verlinde:2007qk} as a means to communicate separated sectors.\footnote{Mixings of hypercharge with closed string RR $U(1)$'s have been discussed in~\cite{Camara:2011jg}. There is no light fields charged under such groups, so we focus in this work exclusively on open string $U(1)$'s.} However, it was shown in~\cite{Shiu:2013wxa} that in such scenarios, charged matter in the hidden sector acquires an electric charge which is necessarily fractional. Given the stringent constraint on the existence of fractionally charged particles~\cite{Langacker:2011db}, mass mixing between hypercharge and hidden $U(1)$'s does not seem an appropriate mechanism to generate interactions between visible and dark matter (DM).\footnote{It is nevertheless an interesting mechanism to mediate supersymmetry breaking to the SM from a hidden sector, as in~\cite{Verlinde:2007qk}, as long as the latter does not contain light fields charged under the hidden $U(1)$.}

We will therefore focus in the following on the role played by massive $U(1)$'s and their mixings as portals into hidden sectors.

\begin{figure}[!htp]
\centering
\includegraphics[width=440pt]{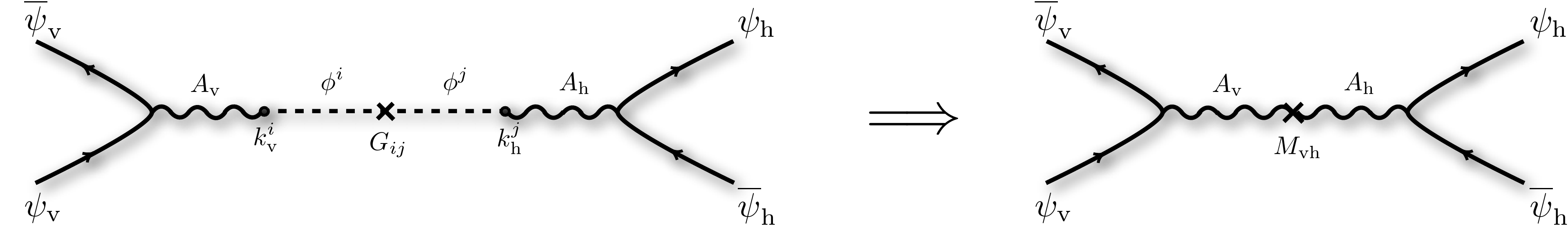}
\caption{\footnotesize Mass mixing of $U(1)$'s as a portal into hidden sectors.}\label{mixingdiagram}
\end{figure}

The crucial point is that $M^2$ can have non diagonal terms that mix $U(1)$'s from different sectors, even when these do not intersect with each other. This can induce a tree-level interaction between visible and hidden matter. The origin of this interaction can be traced back to the St\"uckelberg Lagrangian~\eqref{St\"uckelberg} and is depicted in Figure~\ref{mixingdiagram}. Notice that St\"uckelberg axions involved in the process belong to the closed string sector, and are hence (together with the graviton) natural candidates to connect separated sectors of branes.

The mixing diagrams can be generated either by axions $\phi^i$ that couple both to visible and hidden $U(1)$'s, i.e. for which some $k^i_{\text v}$ and some $k^i_{\text h}$ are non-zero, or by a non-diagonal metric $G_{ij}$ that mixes an axion $\phi^i$ that couples to the visible sector, with an axion $\phi^j$ that couples to the hidden one. Both situations lead to a mass matrix for which off-diagonal terms $M_{\text{vh}}^2\neq 0$.

Given the poor control one usually has over the complex structure moduli space metric $G$ of general CY compactifications, we will focus for the moment on mixings induced by the matrix of axionic charges $K$. Some comments about mixings induced by non-diagonal metrics $G$ will be given in section~\ref{hiddenphotons}.

As mentioned, a tree-level interaction between matter charged under a $U(1)_a\subset U(1)_{\text v}^n$ from the visible gauge group, and matter charged under a $U(1)_{b}\subset U(1)_{\text h}^m$ from the hidden one, will be induced whenever there exists an axion $\phi^i$ for which $k_a^i$ and $k_{b}^i$ are both non-zero. In effect, these generate non-diagonal terms in the matrix of $U(1)$ masses. Since these numbers are (half-)integers one expects the induced mixing effect to be large. 

Recall from~\eqref{masses} that the axionic charges $k^i_a$ are given by the topological numbers
\begin{equation}
k_a^i=N_a \, r_a^i=N_a\, [\alpha^i]\cdot[\Pi_a]\,,
\end{equation}
where $N_a$ is the rank of the gauge group $U(N_a)\supset U(1)_a$, and $r_a^i$ is the number of times the stack wraps around the odd cycle $[\beta_i]$. On the other hand, recall from~\eqref{intersections} that matter charged simultaneously under $U(1)_a$ and $U(1)_b$ arises from the intersections 
\begin{equation}
[\Pi_a]\cdot[\Pi_b]=s_{ai}\,r_b^i- r_a^{\,i}\,s_{bi}\,,~~~~[\Pi_a]\cdot[\Pi_b]'=-s_{ai}\,r_b^i- r_a^{\,i}\,s_{bi}\,.
\end{equation}
Given the fact that both wrapping numbers around even ($s_{ai}$) and around odd ($r_a^{\,i}$) cycles enter these intersections, it is easy to construct models in which two branes do not intersect at all (i.e. $[\Pi_a]\cdot[\Pi_b]=[\Pi_a]\cdot[\Pi_b]'=0$), but there is nevertheless an axion with non-zero charges under both $U(1)_a$ and $U(1)_b$.

\subsubsection*{A toy model}
As a simple example, consider two $U(1)$ branes `v' and `h' that wrap the following cycles:
\begin{equation}
[\Pi_{\text v}]=[\alpha^0]+n[\beta_2]\,,~~~~~~~
[\Pi_{\text h}]=[\alpha^1]+[\beta_1]+m[\beta_2]
\end{equation}
It is straightforward to see that $[\Pi_\text{v}]\cdot[\Pi_{\text h}]=[\Pi_{\text v}]\cdot[\Pi_{\text h}]'=0$, so the branes do not intersect and can belong to separated sectors. Nevertheless, there is a RR axion $\phi^2=\int_{\alpha^2}C_3$ that has charges $k_{\text v}^2=n$ and $k_{\text h}^2=m$. One can hence construct the following diagram that connects both sectors
\begin{figure*}[h!tp]
\centering
\includegraphics[width=160pt]{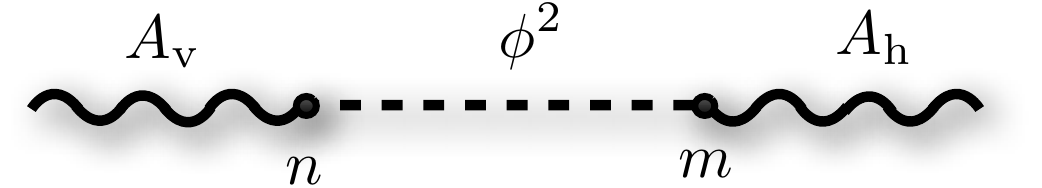}
\end{figure*}

At the level of the effective Lagrangian, the effect is a non-diagonal mass matrix for the $U(1)$ gauge bosons. As mentioned above, we neglect loop-suppressed kinetic mixing terms, so we can write the Lagrangian as
\begin{equation}
\LL=-\frac{1}{4}(F_{\text v} ~ F_{\text h})
\left(
\begin{array}{cc}
g_{\text v}^{-2}&0\\
0&g_{\text h}^{-2}
\end{array}\right)
\left(\begin{array}{c}
F_{\text v}\\
F_{\text h}
\end{array}
\right)
-\frac{1}{2}M_s^2\, (A_{\text v}~A_{\text h})
\left(
\begin{array}{cc}
0& n\\
1&m
\end{array}\right)\cdot G\cdot
\left(
\begin{array}{cc}
0& 1\\
n&m
\end{array}\right)
\left(\begin{array}{c}
A_{\text v}\\
A_{\text h}
\end{array}
\right)
\end{equation}
We can set the canonical kinetic term by rescaling $A_{a}\to g_a A_a$. By diagonalizing the resulting mass matrix, we can read off the physical $Z'$ eigenstates, which will be linear combinations of $A_{\text v}$ and $A_{\text h}$. In order to simplify the results for this toy model, and to illustrate clearly the mixing induced by the non-diagonal $K$ matrix, we will set $G=1$, $g_{\text v}=g_{\text h}\equiv g$ and $n=m=1$. In this case, the physical $Z'$ bosons (those with canonical kinetic term and diagonal mass matrix) read
\begin{eqnarray}
Z'&\propto &(1+\sqrt{5})\,A_{\text v}-2\,A_{\text h}\,\,;\qquad\qquad M_{Z'}^2=\frac{1}{2}(3-\sqrt{5})\,g^2 M_s^2\\
Z''&\propto &(1-\sqrt{5})\,A_{\text v}-2\,A_{\text h}\,\,;\qquad\qquad M_{Z''}^2=\frac{1}{2}(3+\sqrt{5})\,g^2 M_s^2\,.
\end{eqnarray}

The crucial point is that matter currents $J^{\text{(v)}}$, and $J^{\text{(h)}}$, that originally coupled to $U(1)_{\text v}$, and $U(1)_{\text h}$, respectively; will both couple after diagonalization to both $Z'$ and $Z''$:
\begin{eqnarray}
\LL_{\text{int}}&=&q_{\text v} \,A_{\text v} \, J^{\text{(v)}}+q_{\text h}\,A_{\text h} J^{\text{(h)}}\nonumber\\
&=& g'\,Z'\left((1+\sqrt{5})\,q_{\text v}\,J^{(\text v)}-2\, q_{\text h} \,J^{\text{(h}}\right)
+ g''\,Z''\left((1-\sqrt{5})\,q_{\text v}\,J^{(\text v)}-2\, q_{\text h} \,J^{\text{(h}}\right)\,,
\end{eqnarray}
where we have defined the couplings $(g')^2\equiv\frac{g^2}{2(5+\sqrt{5})}$ and $(g'')^2\equiv\frac{g^2}{2(5-\sqrt{5})}$. Clearly, these $Z'$ and $Z''$ bosons connect both sectors.

\subsection{Semi-realistic models}\label{model}
Let us now implement this mechanism in a more realistic setup, in which the visible sector is just the one we described in section~\ref{Madrid}, whose wrapping numbers we rewrite here for completeness:
\begin{align}
[\Pi _{a}^{(\text v)}]&=n_{a}[\alpha ^{0}]+\frac{1}{2}[\alpha ^{1}]+[\beta _{2}]+\frac{n_{a}}{2}[\beta _{3}],\nonumber\\
[\Pi _{b}^{(\text v)}]&=n_{b}[\alpha ^{0}]-\frac{3}{2}[\alpha ^{2}]-[\beta _{1}]+\frac{3n_{b}}{2}[\beta _{3}],\nonumber\\
[\Pi _{c}^{(\text v)}]&=3[\alpha ^{2}]+n_{c}[\beta _{3}],\nonumber\\
[\Pi _{d}^{(\text v)}]&=n_{d}[\alpha ^{0}]-\frac{3}{2}[\alpha ^{1}]-[\beta _{2}]+\frac{3n_{d}}{2}[\beta _{3}],\label{IMRwrappings}
\end{align}
where a label `(v)' has been added to stress that these belong to the visible sector. We want to add a hidden sector to this scenario whose branes do not intersect with the SM ones, but whose $U(1)$ bosons have mixed mass terms with the visible ones.

\subsubsection*{Models with large $h_{2,1}$}
This can be easily done in setups in which the compactification space has enough independent cycles, i.e. where the Hodge number $h_{2,1}$ is large enough. For example, consider the case $h_{2,1}=5$, and modify the wrapping numbers~\eqref{IMRwrappings} as\footnote{Notice that, with this choice, the condition for hypercharge $Y=1/6(Q_a-3Q_b+3Q_d)$ to be massless, eq.~\eqref{masslesshyper} is not modified.}
\begin{equation}
[\Pi_{a}^{(\text v)}]\to[\Pi^{(\text v)}_a]+p\,[\beta_4]\,,\qquad [\Pi_c^{(\text v)}]\to[\Pi_c^{(\text v)}]-p[\beta_4]\,.
\end{equation}
One can then introduce a brane sector composed of two stacks of $N_e$ and $N_f$ branes along
\begin{equation}
[\Pi^{\text{(h)}}_e]=N_f[\alpha^5]+[\beta_4]\,,\qquad [\Pi_f^{\text{(h)}}]=-N_e[\alpha^5]+[\beta_5]
\end{equation}
The hidden sector contains a gauge group $U(N_e)\times U(N_f)$, and charged chiral matter arising from the intersections $[\Pi_e^{\text{(h)}}]\cdot[\Pi_f^{\text{(h)}}]=-[\Pi_e^{\text{(h)}}]\cdot[\Pi_f^{\text{(h)}}]'=N_f$ and $[\Pi_f^{\text{(h)}}]\cdot[\Pi_f^{\text{(h)}}]'=2N_e$. 

One can easily check that there are no intersections between branes from the hidden and visible sectors. However, the axion $\phi^4=\int_{\alpha^4}C_3$ will couple both to the visible $A^{\text{(v)}}_a$ and $A^{\text{(v)}}_c$ gauge bosons, and to the hidden $A^{\text{(h)}}_e$, and hence generate a mass mixing between them. 

Although the induced mixing is generic and controlled by the topological numbers $K_a^i$, in order to write the full mass matrix for the gauge bosons we would need to know the metric $G$ of the complex structure moduli space of the compactification. Details of the internal geometry would be also needed to analyse other aspects such as supersymmetry conditions or stability of the configuration. That is, we would need to pass from the {\it proto}-model discussion we have presented, to an explicit compactification setup.

Unfortunately, these details are under poor control for generic CY compactifications. Nevertheless, since the mass mixing mixing mechanism relies only on the topology of the internal space, it is expected that these {\it proto}-models find implementations in concrete setups. Furthermore, it is clear that for topologies with larger Hodge numbers, one can generalize our models and include more hidden branes with massive $U(1)$ bosons that mix with the visible ones.

\subsubsection*{An explicit toroidal model}
Even though detailed computations cannot be carried out for generic compactifications, it is possible to implement the St\"uckelberg mechanism even in the simplest toroidal setups discussed in section~\ref{Madrid}, where one can find explicit expressions for the moduli space metric $G$ and other geometric quantities.

The relevant Hodge number of the torus is $h_{2,1}({\bf T}^6)=3$, so there are four axions $\phi^i$ $(i=0,1,2,3)$ that can be swallowed by $U(1)$ bosons via the St\"uckelberg coupling. The visible sector~\eqref{IMRwrappings} already contains four visible $U(1)$'s, out of which three gain a St\"uckelberg mass (all but hypercharge). Hence, in ${\bf T}^6$ there is only  space for one more massive $U(1)$, which we will try to locate in a hidden sector. 

Let us consider an additional brane stack that does not intersect those in eqs.~\eqref{IMRwrappings} (nor their images), and whose world-volume gauge theory contains a $U(1)$ factor that gains a mass by the St\"uckelberg mechanism. One can check that such a brane cannot be added for a generic choice of the parameters $(n_a,n_b,n_c,n_d)$ in~\eqref{IMRwrappings}. In fact, the choices that are compatible with condition~\eqref{masslesshyper} for a massless hypercharge, and with tadpole cancellation conditions are quite restrictive, but not inexistent. One such choice is given by $(n_a,n_b,n_c,n_d)=(1,0,-4,-3)$, for which eqs.~\eqref{IMRwrappings} reads
\begin{align}
[\Pi _{a}^{(\text v)}]&=[\alpha ^{0}]+\frac{1}{2}[\alpha ^{1}]+[\beta _{2}]+\frac{1}{2}[\beta _{3}],\nonumber\\
[\Pi _{b}^{(\text v)}]&=-\frac{3}{2}[\alpha ^{2}]-[\beta _{1}],\nonumber\\
[\Pi _{c}^{(\text v)}]&=3[\alpha ^{2}]-4[\beta _{3}],\nonumber\\
[\Pi _{d}^{(\text v)}]&=-3[\alpha ^{0}]-\frac{3}{2}[\alpha ^{1}]-[\beta _{2}]-\frac{9}{2}[\beta _{3}],\nonumber\\
&\rule{5cm}{0.4pt}\nonumber\\
[\Pi^{(\text h)}]&=n_{\text h}[\alpha ^{0}]-p_{\text h}[\beta_0]+2p_{\text h}[\beta_1]+m_{\text h}[\beta _{3}]\,.\label{ourwrappings}
\end{align}
In this scheme we have already added an extra stack of $N_{\text h}$ branes along $[\Pi^{\text{(h)}}]$ giving rise to a gauge group $SU(N_{\text h})\times U(1)_{\text h}$. Its wrapping numbers are determined by three discrete parameters $n_{\text h}$, $p_{\text h}$ and $m_{\text h}$. It can be seen that the corresponding cycle does not intersect those wrapped by branes from the visible sector, so there is no chiral matter charged simultaneously under both the hidden and the SM gauge group. 

Nevertheless, there is chiral matter charged under the hidden gauge group arising from the intersection of $[\Pi^{\text{(h)}}]$ with its orientifold image $[\Pi^{\text{(h)}}]\cdot[\Pi^{\text{(h)}}]'=2n_{\text h} p_{\text h}$, that gives rise to chiral fermions charged under $SU(N_{\text h})\times U(1)_{\text h}$. In particular, there are $p_{\text h}(n_{\text h}-4)$ chiral fermions in the symmetric $\tiny\yng(2)_{a,+2}$ and $p_{\text h}(n_{\text h}+4)$ in the antisymmetric $\tiny\yng(1,1)_{a,+2}$ representation. 

For the wrapping numbers~\eqref{ourwrappings}, tadpole cancellation imposes $N_h n_h=16$, so we can choose the rank of the hidden group to be $N_h=1,2,4,8,16$, and correspondingly $n_h=16,8,4,2,1$. On the other hand $p_h$ is an unconstrained integer number, while $m_h$ is a discrete parameter that, because of the tilting of the third torus, must be an integer if $n_h$ is even, and a half-integer if $n_h$ is odd.

It is worth noting here that, unlike the visible cycles, $[\Pi^{\text{(h)}}]$ is not factorizable, i.e. its homology cannot be written in terms of wrapping numbers $(n_i,m_i)$ along each ${\bf T}^2$. Nevertheless, these non-factorizable cycles can always be expressed as the linear combination of two factorizable ones, and can be thought of as the result of their recombination~\cite{Rabadan:2001mt}.

Let us now discuss the masses and mixings of the $U(1)$ gauge bosons of the setup. It is clear already from~\eqref{ourwrappings} that the axions $\phi^1=\int_{\alpha^1}C_3$ and $\phi^3=\int_{\alpha^3}C_3$ couple both to the visible and the hidden sector. In fact, the matrix $K$ of axionic charges reads
\begin{equation}\label{Ktorus}
K=\left(
\begin{array}{ccccc}
0&0&0&0&-N_h p_h\\
0&-2&0&0&2N_h p_h\\
3&0&0&-1&0\\
3/2&0&-4&-9/2&N_h m_h
\end{array}
\right)
\end{equation}
This generically generates a highly non-diagonal mass matrix $M^2=K^{\,T}\cdot G\cdot K$. Indeed, for the case of the torus, the complex structure moduli space metric is known to be diagonal at tree-level. It takes the form (see e.g.~\cite{Ibanez:2012zz,Grimm:2005fa})
\begin{equation}\label{Gtorus}
G_{ij}=\frac{\delta_{ij}}{4 \kappa_4^2 ({\text{Re }} U^i)^2}
\end{equation}
where the complex structure moduli $U^i$ defined in~\eqref{csmoduli} can be related to the radii of the ${\bf T}^6$, and 
\begin{equation}
\kappa_4^2=\frac{4\pi\alpha'}{\sqrt{(U_0+U_0^*)(U_1+U_1^*)(U_2+U_2^*)(U_3+U_3^*)}}
\end{equation}
Finally, the gauge couplings of the $U(1)$ bosons are encoded in their kinetic matrix $f$, which at tree level is diagonal and reads $f_{ab}=\delta_{ab} g_a^{-2}$. For a brane wrapping the cycle $[\Pi_a]=a_i[\alpha^i]+b^j[\beta_j]$, the couplings are
\begin{equation}\label{gtorus}
f_{aa}=\frac{1}{g_a^2}=\frac{1}{(2\pi)^4}\left\{\left[a_0 {\mathrm{Re}}(U^0)-a_i{\mathrm{Re}}(U^i)\right]^2+
(4\pi\alpha')^2\left[b^0 G_{00}{\mathrm{Re}}(U^0)-b^i G_{ii}{\mathrm{Re}}(U^ i)
\right]^2\right\}^{1/2}
\end{equation}
which basically correspond to the volumes of the branes in the internal space.
With these ingredients, we can compute the mass matrix of $U(1)$ gauge bosons, and by finding it's eigenvalues, read off the `physical' $Z'$ bosons (those with canonical kinetic term and diagonal mass matrix), and their couplings to matter. We will perform this analysis explicitly in the following sections in order to exemplify the generic St\"uckelberg portal we have described.

\subsection{Supersymmetry and stability}
Let us first, however, address the important question of whether the configurations we discuss are supersymmetric, and whether they are stable or not. Some MSSM-like setups have been constructed in the literature (see \cite{Cvetic:2001tj,Cvetic:2001nr,Blumenhagen:2002gw,Honecker:2003vq,Honecker:2004kb} for early toroidal constructions and~\cite{Blumenhagen:2002vp,Palti:2009bt} for their implementation in more general CY), but the simple examples we have considered in sections \ref{Madrid} and \ref{model} are non-supersymmetric. In such cases, scalar fields arising at the brane intersections may be massive, massless, or tachyonic, depending on geometrical details of the compactification, and in particular on the values at which the complex structure moduli are stabilized.

Given the lack of control on the geometrical details of generic CY compactifications, it is hard to address such issues of stability in full generality.
For the toroidal models introduced in section \ref{Madrid}, however, it was shown in~\cite{Ibanez:2001nd} that there are regions in the complex structure parameter space for which all the scalars of the system are massive and the configuration is stable. This was a crucial check for the validity of the models.

When studying the generalized setups showing the St\"uckelberg portal of section \ref{model}, one would need to analyze whether the branes of the hidden sector are stable or not, i.e. whether the scalar fields that appear at their intersections are tachyonic or not. A tachyon in the hidden sector, however, would not invalidate the model, but would rather indicate that the hidden branes tend to recombine with each other. This recombination has a low energy interpretation in terms of a Higgs mechanism by which chiral matter living in the hidden sector intersections (which are good DM candidates) would acquire a mass. Hence, the possible appearance of open string tachyons in the hidden sector could be an advantage rather than a problem. In the particular toroidal models of eq.~\eqref{ourwrappings}, the tachyons could appear at the intersections of $[\Pi^{(\text{h})}]$ with its orientifold image $[\Pi^{(\text{h})}]'$, and would trigger a recombination of branes into a stack of branes aligned with the orientifold plane.

 Since we are concerned in this paper with the general mechanism of the St\"uckelberg portal into the hidden sector, and not so much on the particular dynamics of the hidden sector, we will not address the issue of stability further and proceed directly to the study of the $Z'$ bosons in our constructions.

\section{Phenomenology of the St\"uckelberg portal}\label{sec:pheno}
Let us now study some of the phenomenological properties of the St\"uckelberg portal scenario presented in the previous section. From a low energy perspective, these models amount to an extension of the SM by a number of massive $Z'$ bosons that mediate interactions of visible matter fields with a hidden sector that has its own gauge group and matter content.

The `physical' $Z'$ eigenstates, i.e. those that propagate without mixing, are obtained by diagonalizing simultaneously the kinetic and mass matrix of the $U(1)$ bosons. After doing so, one can read off their masses and couplings to matter fields, and these determine the main phenomenological features of the scenario. In particular, the lightest $Z'$ boson may lead to very interesting astrophysical and particle collider signatures if its mass is light enough. 

In this section, we study the role played by this lightest $Z'$ eigenstate in the communication of the visible and the hidden sectors. We will often rely on the toroidal models described in section~\ref{model} in order to illustrate generic features of this St\"uckelberg portal scenario with explicit computations that depend on a small number of parameters.

\subsection{Mass eigenstates}
At tree level, the kinetic and mass matrices of the $U(1)$ bosons are $f_{ab}=\delta_{ab}g_a^{-2}$ and $M^2=K^{\,T}\cdot G\cdot K$, where $K$, $G$ and $g_a$ were given for our particular toroidal model in equations~\eqref{Ktorus}, \eqref{Gtorus} and \eqref{gtorus}, respectively. The model depends only on a few discrete parameters, namely $N_h$, $n_h$, $p_h$ and $m_h$ (of which $N_h$ and $n_h$ are related by the tadpole cancellation condition $N_h n_h=16$); as well as on the particular values at which the four complex structure moduli $u^i$ are stabilized.\footnote{We will not discuss here the particular mechanism by which the moduli are stabilized.} 

We will focus here on the case $N_h=1$, $n_h=16$, for which the hidden sector group is simply $U(1)$ and there are 12  chiral hidden fermions with charge +2, and will take $m_h$ as a free integer parameter.\footnote{For this case in which the hidden group is abelian, there is not such thing as an asymmetric representation, so the corresponding part of the spectrum is absent after the orientifold projection.}  We also set $p_h=1$ for illustration. Other choices can be analyzed in a similar way. 

The continuous parameters $u^i={\text{Re}}\,(U^i)$ are related to the gauge coupling constants $g_a$ by~\eqref{gtorus}, and three of them can be fixed by requiring that the couplings of the SM are reproduced. After normalizing as usual the quadratic Casimir in the fundamental representation to be 1/2, we obtain the relations~\cite{Ghilencea:2002da}
\begin{equation}\label{couplings}
g_a^2=\frac{g_{\text{QCD}}^2}{6}\,,~~~~~g_b^2=\frac{g_L^2}{4}\,;
\end{equation}
and from~\eqref{hyper} we get
\begin{equation}\label{hypercoupling}
\frac{1}{g_Y^2}=\frac{1}{36g_a^2}+\frac{1}{4g_c^2}+\frac{1}{4g_d^2}\,.
\end{equation}
Equations~\eqref{couplings} fix the values of $g_a$ and $g_b$, while~\eqref{hypercoupling} constrains $g_c$ and $g_d$. Following the analysis of~\cite{Ghilencea:2002da} we will write our results in terms of $g_d/g_c$, which we take as a free parameter that must lie in the range $g_c/g_d\in[0.84, 2.19]$, in which the real complex structure moduli $u^i$ are all positive (c.f. equation~\eqref{gtorus}), and equations~\eqref{couplings} and~\eqref{hypercoupling} can be satisfied.\footnote{The SM couplings should be evaluated at the string scale using the renormalization group equations. For reference we use their values at 100 TeV for explicit computations. As we will momentarily see, it is around this scale that the phenomenology of the $Z'$ bosons becomes more interesting. Given the logarithmic running of the coupling constants, we do not expect drastic deviations from our results for other string scales not far from this reference value.}

In order to find the physical $Z'$ eigenstates of the model we have to go first to a basis in the space of $U(1)$ gauge bosons in which their kinetic term is canonical. This is done by rescaling the gauge bosons as $A_a\to g_a A_a$, or in matrix notation $\vec{A}\to\hat{g}\cdot\vec{A}$, where $\hat{g}\equiv \text{diag}(g_a, g_b,\ldots, g_N)$. After doing so, the physical $Z'$ bosons will correspond to the eigenvectors of the resulting mass matrix 
\begin{equation}
\tilde{M}^2=\hat{g}\cdot K^{\,T} \cdot G\cdot K\cdot\hat{g}\,,
\end{equation}
with masses corresponding to its eigenvalues $m_i^2$.

Given a set of normalized eigenvectors $\vec{v}_i$ of the mass matrix $\tilde{M}$, we can identify the physical state $Z'_i$ associated to each of them with the linear combinations of the original $U(1)$ gauge bosons given by
\begin{equation}\label{trafo}
Z'_i=\sum_a \frac{1}{g_a}\, (\vec{v}_i)^{a}\,  \vec{A}_a\,.
\end{equation}
Hypercharge can be identified with the eigenvector $Z'_0\equiv A_Y=\frac{1}{6g_Y}(A_a-3A_c+3A_d)$ that has zero eigenvalue.

Plugging transformation~\eqref{trafo} back into the original lagrangian~\eqref{lag}, one obtains
\begin{eqnarray}
\LL&=&-\frac{1}{4}\vec{F}^{\,T}\cdot f\cdot\vec{F}-\frac{1}{2} \vec{A}^{\,T}\cdot M^2\cdot\vec{A}+\sum_r\vec{q}_{r}^{\,T}\cdot\vec{A}\,J^{\,r}\nonumber\\
&=&\sum_i\left(-\frac{1}{4}{F'}_i^{2}-\frac{1}{2}\,m_i^2\,{Z'}_i^2+\sum_r (\vec{q}^{\,T}_r\cdot\hat{g}\cdot\vec{v}_i) \, Z'_i J^{\,r}\right)\,,
\end{eqnarray}
from which we can read the physical gauge boson masses $m_i$ and their couplings to the matter currents $J^{\,r}$, which we denote $g'_{ir}\equiv (\vec{q}^{\,T}_r\cdot\hat{g}\cdot\vec{v}_i)$.

\begin{figure}[h!tb]
\begin{center}
  \includegraphics[scale=0.5]{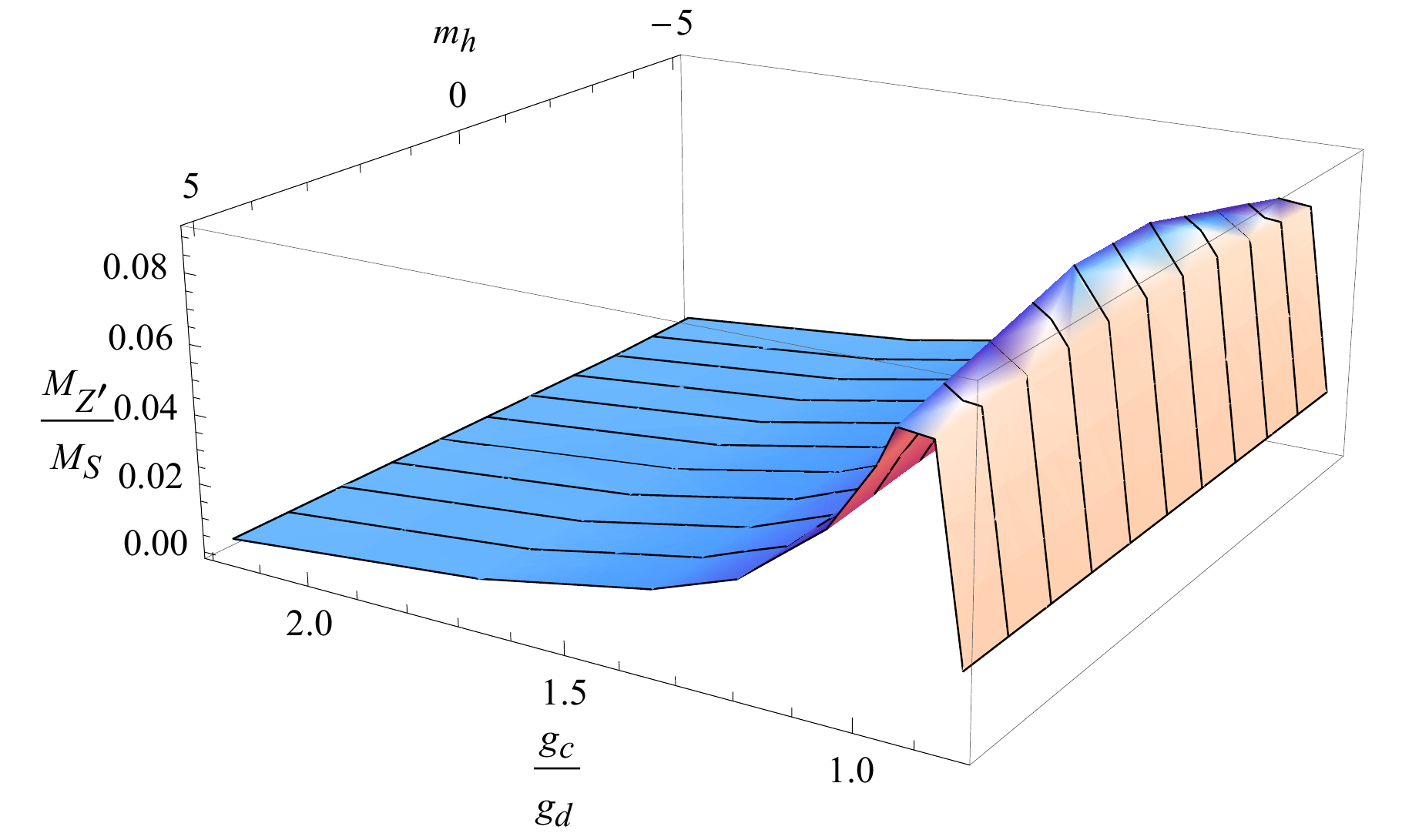}
  \caption{Mass of the lightest $Z'$ boson in terms of the string scale $M_s$, as a function of the continuous parameter $g_c/g_d$ and the discrete variable $m_h$, for the toroidal model with wrapping numbers~\eqref{ourwrappings}.}
  \label{massplot}
\end{center}
\end{figure}

Both the mass eigenvalues $m_i$ and eigenvectors $\vec{v}_i$ (and hence the couplings $g'_{ir}$) can be computed numerically. In figure~\ref{massplot} we show the mass of the lightest (non massless) $Z'$ boson (corresponding to the smallest non-zero eigenvalue of $\tilde{M}$) in terms of the string scale mass $M_s=1/\sqrt{\alpha'}$, as a function of the free parameters $g_c/g_d$ and $m_h$.

\begin{figure}[h!tb]
\begin{center}
  \includegraphics[scale=0.33]{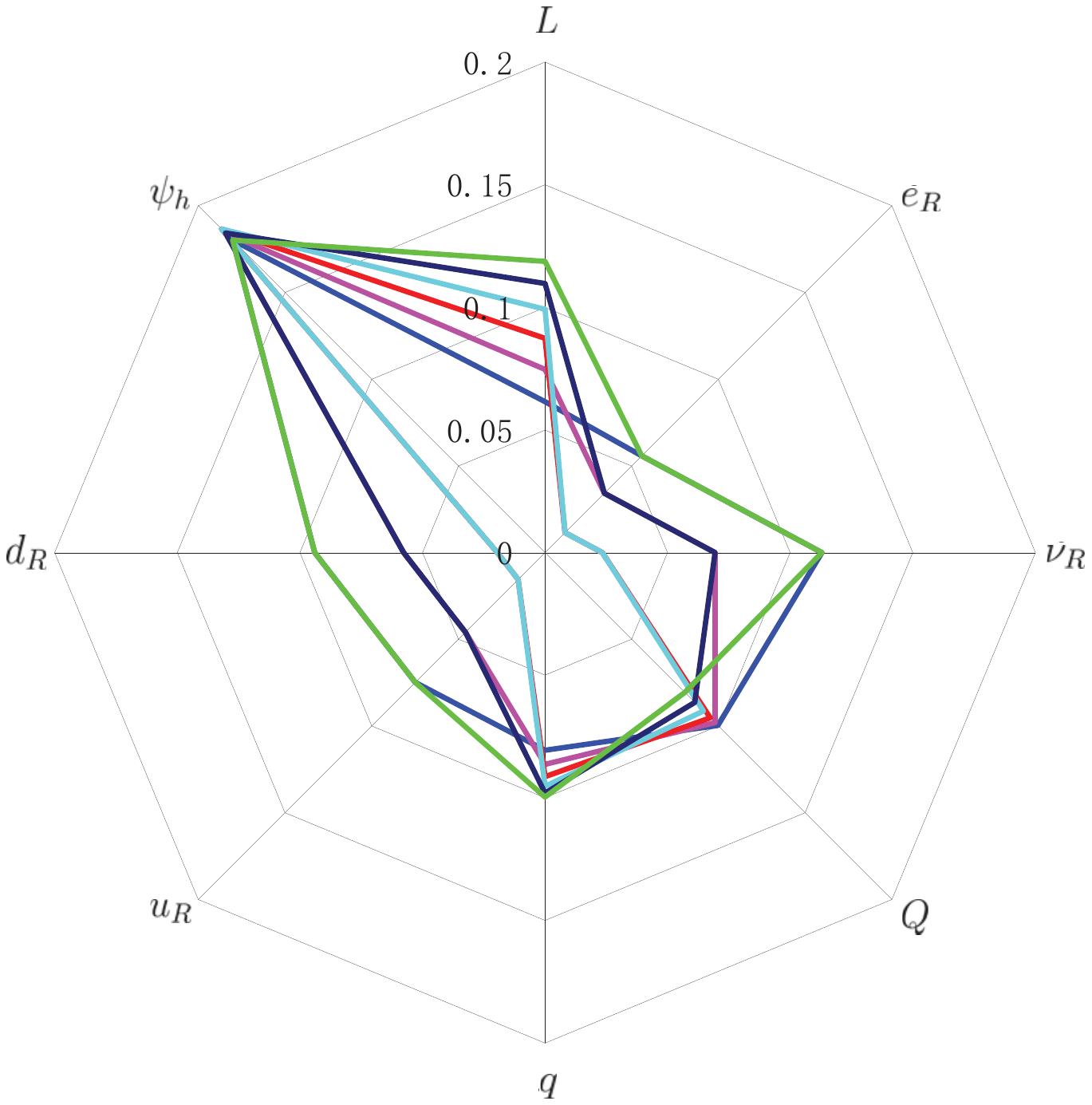}
  \includegraphics[scale=0.33]{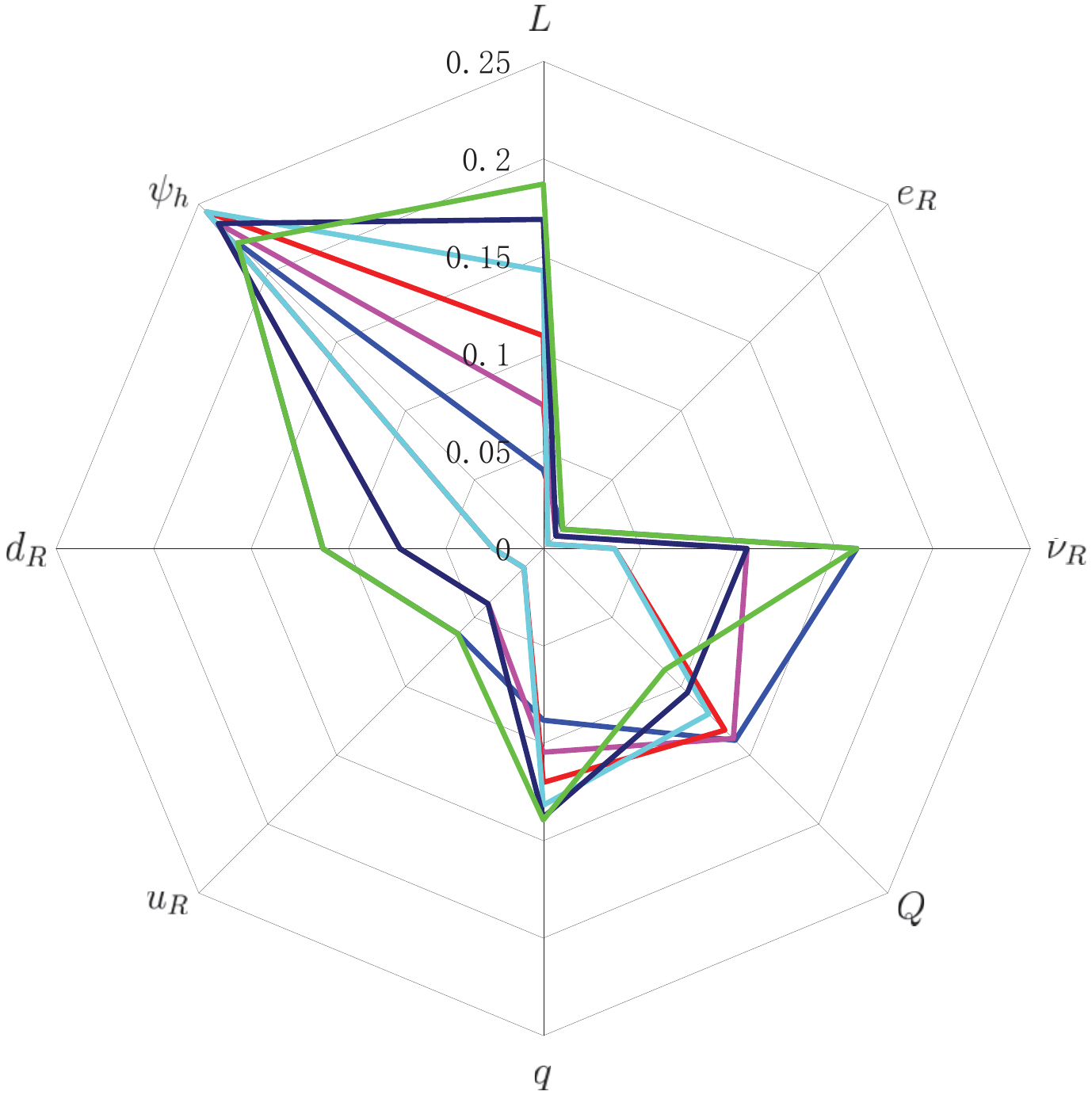}
  \includegraphics[scale=0.33]{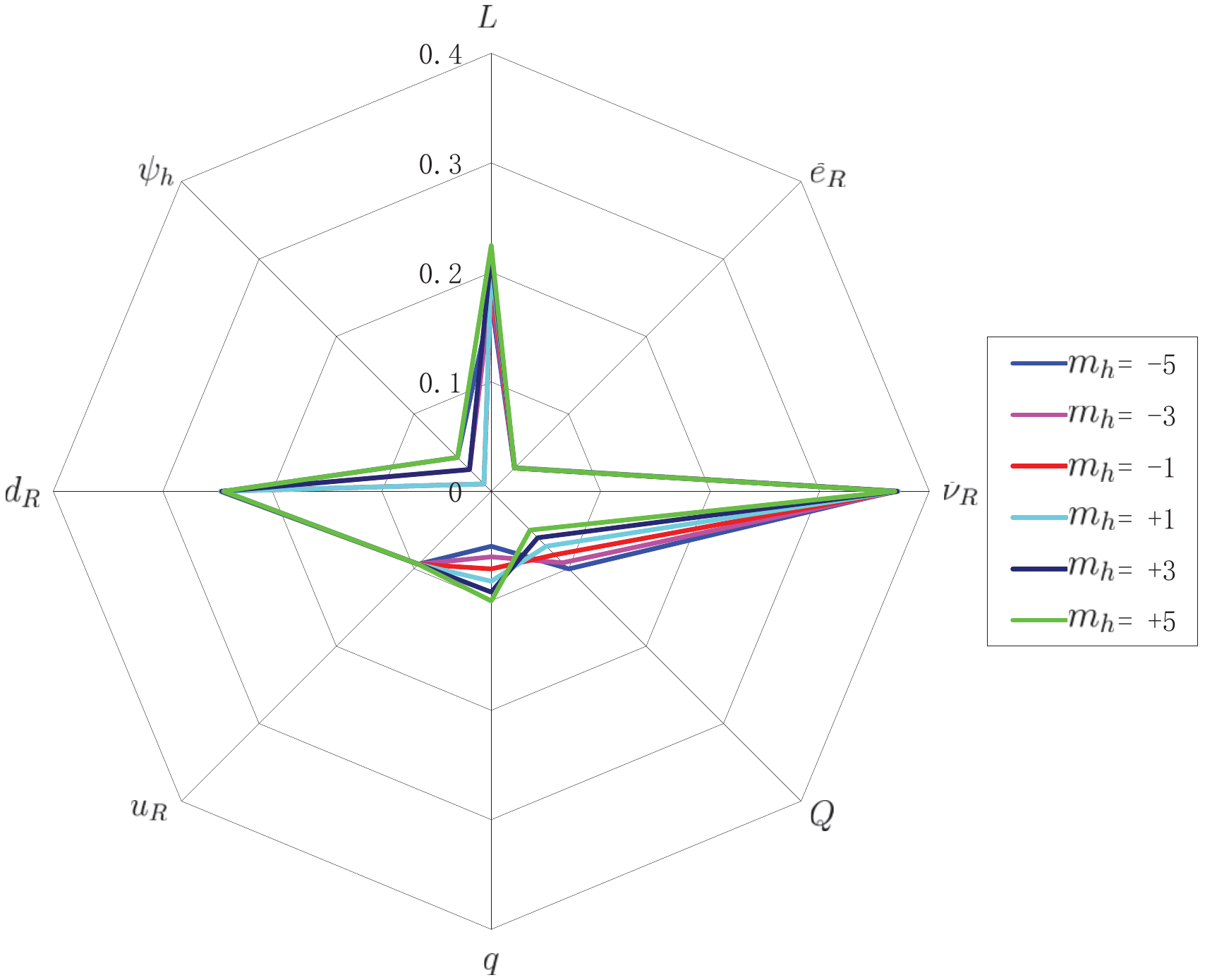}
  \caption{A depiction of the coupling strength of the lightest $Z'$ to visible and hidden sector fields,
  for $g_{c}/g_{d}=1.68$ (left), $g_{c}/g_{d}=1.06$ (middle), and $g_{c}/g_{d}=0.89$ (right).
  The parameter
  $m_h$ is chosen to be $\pm1,\pm3,\pm5$.}
  \label{couplingsplot}
\end{center}
\end{figure}

Analogously, we present in figure~\ref{couplingsplot} the couplings of the lightest $Z'$ boson to the matter fields of the visible sector (the SM matter fields, whose charges were presented in table~\ref{IMR}) as well as the hidden ones for several values of the free parameters $g_c/g_d$ and $m_h$.

\subsection{EWSB and $Z-Z'$ mixing}
So far, we have discussed the mixing of $U(1)$ bosons before Electro Weak Symmetry Breaking (EWSB). At energies lower than the EW scale, however, it is important to take into account the Higgs vev $v$, which gives a mass to a certain linear combination of $U(1)$ bosons and the third component of $SU(2)_L$. As discussed in section~\ref{sec:higgs}, this effect can be incorporated into our discussion, by simply including the neutral component $A^L_3$ of $SU(2)_L$ among the $U(1)$ vector bosons $\vec{A}$, and correspondingly extending the kinetic and mass matrices of the gauge bosons to include the $SU(2)_L$ coupling $g_L$, and the Higgs field's vev and charge vector:
\begin{align}
\vec{A}\to {\vec{A}}_{EW}&=\left(\begin{array}{c}A^L_3\\ \vec{A}\end{array}\right)&\\
f\to f_{EW}= \left(\begin{array}{cc} g_L^{-2} & 0\\ 0&f \end{array}\right) \,,~~~ 
K\to K_{EW}&=\left(\begin{array}{cc} -1 & \vec{q}^{\,T}_H \\ \vec{0} & K \end{array}\right)\,,~~~
G\to G_{EW}=\left(\begin{array}{cc} v^2 &0 \\ 0 & G \end{array}\right)\,,&\nonumber
\end{align}
where $\vec{q}_H$ denotes the vector of charges of the Higgs field under the $U(1)$ gauge bosons $\vec{A}$. 

After EWSB, the photon will be the only massless gauge boson in the system and will take the usual SM form
\begin{equation}\label{photon}
A_\gamma = e \left(\frac{A_3^L}{g_L} + \frac{A_Y}{g_Y}\right)
\end{equation}
where $e=\frac{g_L g_Y}{\sqrt{g_L^2+g_Y^2}}$ is the usual electric coupling constant.\footnote{The gauge coupling constants enter eq.~\eqref{photon} after the rescaling $A_a\to g_a A_a$ to set a canonical kinetic term $f\to 1$, see~\cite{Ghilencea:2002da,Shiu:2013wxa} for more details.} 

The lightest massive eigenstate is identified with the SM $Z$ boson. However, the presence of other massive $Z'$ bosons and the non-diagonal mass matrix $M_{EW}=K_{EW}^{\,T}\cdot G_{EW}\cdot K_{EW}$ results in a small $Z-Z'$ mixing  and hence in deviations from the SM predictions for the EW parameters. For example, the $Z$ mass receives corrections of the form~\cite{Ghilencea:2002da}
\begin{equation}
M_Z=\underbrace{\frac{v}{2}\sqrt{g_L^2+g_Y^2}}_{\equiv M_{Z,0}}\,+\,{\cal O}\left(\frac{M_{Z,0}}{M_{Z'}}\right)\,,
\end{equation}
where $M_{Z,0}$ is the value predicted by the SM. Ultimately, the $Z-Z'$ mixing is due to the fact that the Higgs boson couples not only to hypercharge, but also to the massive $Z'$ bosons of the system.

It is worth noting that in many previous phenomenological models with extra $U(1)$'s, $Z-Z'$ mixing prevented strong couplings of the SM particles to the $Z'$ boson. In our setups, however, these couplings can be rather large while evading constraints from EW precision measurements. 

\subsection{Experimental bounds and implications}\label{sec:exp}

From figure~\ref{couplingsplot} we can see that the $Z'$ bosons couple both to SM and hidden matter with significant strength. 
Current experiments set strong constraints on the mass and couplings of such bosons and on their mixing with the SM $Z$. We discuss briefly some of these constraints and prospects for detection of such heavy $Z'$ bosons.
For more details on the phenomenology of heavy $Z'$ scenarios see~\cite{Rizzo:2006nw, Langacker:2008yv} and references therein.

LEP II and LHC both put stringent bounds on the properties of $Z'$ bosons. At LHC, such fields could appear as resonances at the $Z'$ mass and be detected  through Drell-Yan processes $pp\to Z' \to l^+l^-$ where $l=e,\mu$~\cite{Accomando:2010fz,Chatrchyan:2012it}, or by examining their dijet resonances~\cite{Chatrchyan:2013qha}. Constraints from $e^+e^-$ colliders come from precision measurements at the $Z$ pole, and from resonance productions at $e^+e^-\to l^+l^-$ processes~\cite{LEP:2003aa,Carena:2004xs}. Although the results depend on the particular models, i.e. on the couplings of the $Z'$ under consideration, one can generically take $M_{Z'}\gtrsim 2$ TeV as a reasonable bound.\footnote{It should be noted that most analysis are performed assuming exclusive decays of $Z'$ into SM particles. In our models, decays into hidden particles are also possible, resulting in an increase of the $Z'$ decay width. This may weaken the bounds, as well as the sensitivity to discover these $Z'$ bosons in colliders.}

Constraints on the coupling of the $Z'$ boson to muons also come from contributions to the muon anomalous magnetic moment, which reads
\begin{equation}
\Delta(g_{\mu}-2)=-\frac{m_{\mu}^{2}}{6\pi^{2}M_{Z'}^{2}}\left(g_{L}^{\mu\,2}-3g_{L}^\mu g_{R}^\mu+g_{R}^{\mu\,2}\right)\,,
\end{equation}
where $g_{L}^\mu$ and $g_{R}^\mu$ are the couplings of $Z'$ to the left-handed and
the right-handed muon, respectively. The $Z'$ contribution should not exceed the commonly adopted experimental ($4\sigma$) deviation of $\Delta\big((g_{\mu}-2)/2\big)=(3.0\pm0.8)\times10^{-9}$~\cite{Miller:2007kk}.
A simple calculation shows that if $M_{Z'}\gtrsim2~{\rm {TeV}}$ the constraint on the couplings $g_{L}^\mu$ and $g_{R}^\mu$ is very weak.

If the mass of the $Z'$ is within the reach of LHC or future hadron colliders with higher energies,
$Z'$ would show up in the resonant production in $pp$ collision at the $Z'$ mass peak.
As one can see from Fig.~\ref{couplingsplot}, the couplings between the $Z'$ and the SM fermions,
and in particular the couplings of $Z'$ to left- and right-handed components of the same fields are generically different.
One could use the branching ratios of the different $Z'$ decay channels into SM fermion pairs,
and also the forward-backward asymmetry in observed channels as distinguishing features of this type of scenario. 
Furthermore, massive gauge bosons from anomalous $U(1)$ symmetries may present trilinear couplings to SM gauge bosons and result in exotic decay channels into $ZZ$, $WW$ and $Z\gamma$~\cite{Anastasopoulos:2006cz,Coriano:2005js,Armillis:2007tb,Berenstein:2006pk,Kumar:2007zza,Berenstein:2008xg} (astrophysical signatures arising from such anomaly related couplings have been studied in~\cite{Dudas:2009uq,Mambrini:2009ad,Dudas:2013sia}).

In order to test whether these $Z'$ bosons do indeed couple to a hidden sector and realize the St\"uckelberg portal, one would have to find evidence of their interactions with hidden matter. At colliders, these would result in processes with missing energy from decays $Z'\to \overline{\psi}_h\psi_h$. On the other hand, if matter fields from the hidden sector realize DM (see next subsection), they could also be directly observed via their elastic scattering with nucleons through the $t$-channel exchange of the $Z'$ boson at DM detectors. See~\cite{Alves:2013tqa,Arcadi:2013qia} for a recent analysis of these experimental signatures.

The spin-independent DM-nucleon (target-independent) cross section,
which receives bounds from DM direct detection experiments,
can be approximately written as
\begin{equation}
\sigma_{{\rm SI}}\approx\frac{4}{\pi}\mu_{n}^{2}f_{n}^{2}\,,
\end{equation}
where $\mu_{n} \sim 1$~GeV is the reduced mass of the dark particle and the nucleon (assuming $m_{DM}\gg$~GeV), and
$f_n \sim g^2 / M_{Z'}^2$ is their effective spin-independent coupling.
For a $Z'$ gauge boson with a 2 TeV mass and matter couplings of order $g\approx0.1$,
one obtains $\sigma_{{\rm SI}}\sim10^{-44}\,{\rm cm}^{2}$,
which is just on the edge of the recently released LUX data~\cite{Akerib:2013tjd} for $\sim 1$~TeV DM masses.

\subsection{Dark matter stability and  relic density}

In our models, matter fields in the
hidden sector are natural DM candidates. Our main goal in this work is to study the interactions of these fields with the visible sector through exchange of $Z'$ bosons, while we are not so much concerned with the particular features of the hidden sector, which could take widely different forms in different D-brane constructions. Nevertheless, there are a few rather generic features of St\"uckelberg portal models that we would like to discuss here.

A simple consequence of the presence of $U(1)_h$ symmetries under which hidden fermions are charged is that the lightest of these matter fields would be stable. Despite gaining a mass through the St\"uckelberg mechanism, the $U(1)$ symmetries remain unbroken at the perturbative level in the low energy effective theory~\cite{Ibanez:1999it}, and can easily guarantee stability of DM in our scenarios. Nevertheless, there are D-brane instanton effects that break these symmetries at the non-perturbative level ({\it c.f.} section~\ref{sec:non-perturbative}), which could trigger instabilities of DM. In cases in which such effects were too large to allow for a sufficient DM lifetime, one could easily consider the implementation of exact discrete symmetries that forbid some of such dangerous couplings. 

The implementation of such discrete symmetries in intersecting brane models has been studied in~\cite{BerasaluceGonzalez:2011wy} (see also~\cite{Camara:2011jg,Anastasopoulos:2012zu,Honecker:2013hda,Marchesano:2013ega} ), where the focus was put into discrete symmetries (such as R-parity) that prevent proton decay. It is easy to apply a similar analysis to our scenarios to obtain conditions under which discrete subgroups of the hidden sector abelian symmetries remain exact. In the explicit toroidal models of equation~\eqref{ourwrappings}, one can see that a $\mathbb{Z}_s\subset U(1)_h$ subgroup is not broken by non-perturbative effects, where $s=\mathrm{g.c.d.}(n_h,p_h)$. The key point is that such effects are induced by operators of the form $e^{-U^i}$ which behave as fields with $U(1)_h$ charge $N_h r_h^i$ (c.f. footnone~\ref{footnote} and section~\ref{sec:non-perturbative}). All of these charges are multiple of $s$, so indeed $\mathbb{Z}_s\subset U(1)_h$ would not be broken by such effects.

Besides stability, a major issue in the phenomenology of DM is the analysis of its annihilation rate and relic density. These factors depend drastically on the particular features of the hidden sector (field content, dynamics, etc) and a general analysis cannot be performed at the level of our discussion. Nevertheless, it is worth mentioning here that the annihilation of hidden matter into SM fields through the $Z'$ pole
\begin{equation}
\bar{\psi}_{h}+\psi_{h}\to Z'\to\bar{\psi}_{v}+\psi_{v}\,.
\end{equation}
can be an efficient mechanism to reduce the hidden particle primordial density and to achieve the current value of the DM relic density.

Let us study this process in some more detail. For the following computation, we will assume that DM is composed of  $N_{\psi_h}$ families of massive hidden fermions $\psi_h$, all with the same mass $m_{\psi}$, which are totally stable and annihilate exclusively through the $Z'$ pole. In such a simplified model, one can estimate the resulting relic density as~\cite{Griest:1990kh,Gondolo:1990dk,Nath:1992ty}
\begin{equation}
\Omega_{\psi}h_{0}^{2}=\Omega_{\bar{\psi}}h_{0}^{2} \approx 2.2\times10^{-11}\frac{h(x_{0})}{\sqrt{h(x_f)}}\frac{1}{J(x_{f})}\,.
\end{equation}
Here $x=k_{B}T/m_{\psi}$, and $h(x_{f})$ and $h(x_{0})$
are the entropy degrees of freedom at freeze out and at the current
temperature respectively.  $x_{f}$
is typically of size $\sim1/20$. The function $J(x_f)$ is given by the integral
\begin{equation}
J(x_{f})\equiv\int_{x_{0}}^{x_{f}}\langle\sigma v\rangle\,\mathrm{d}x\,.
\end{equation}
of the total annihilation cross section $\sigma=\sum_{f}\sigma_{\bar{\psi}\psi\to\bar{f}f}$ thermally averaged using the Boltzmann distribution:
\begin{equation}\label{averaged}
\langle\sigma v\rangle= \frac{\displaystyle{\int_0^\infty} dv \, (\sigma v) v^2 \,  e^{-v^2/4x}\,}{\displaystyle{\int_0^\infty} dv \,  v^2 \,  e^{-v^2/4x}} \,.
\end{equation}

Using the Breit-Wigner form,
the partial annihilation
cross sections is
\begin{align}\label{cross}
\sigma_{\bar{\psi}\psi\to\bar{f}f} & =a_{\psi}\left|\left({s-M_{Z'}^{2}+i\Gamma_{Z'}M_{Z'}}\right)\right|^{-2}\,,\\
a_{\psi} & =N_{\psi_{h}}\frac{\beta_{f}(g_{f}g_{\psi})^{2}}{64\pi s\beta_{\psi}}\left[s^{2}(1+\frac{{1}}{3}\beta_{f}^{2}\beta_{\psi}^{2})+4m_{\psi}^{2}(s-2m_{f}^{2})+{4}m_{f}^{2}(s+2M_{\psi}^{2})\right]\,,\nonumber
\end{align}
where $N_{\psi_{h}}$ denotes the multiplicity of the hidden matter fields $\psi$, and $\beta_{f,\psi}=(1-4m_{f,\psi}^{2}/s)^{1/2}$.
$\Gamma_{Z'}$ is the total decay width of $Z'$, i.e., the sum of
the $Z'$ partial decay widths in all the $Z'$ decay channels, which are given in terms of the couplings and masses by
\begin{equation}\label{width}
\Gamma(Z'\to\bar{\psi}\psi)=(g_{L}^{2}+g_{R}^{2})\frac{M_{Z'}}{24\pi}\Big(1+\frac{2m_{\psi}^{2}}{M_{Z'}^{2}}\Big)\sqrt{1-\frac{4m_{\psi}^{2}}{M_{Z'}^{2}}}\ \Theta\left(M_{Z'}-2m_{\psi}\right)\,.
\end{equation}

\begin{figure}[t!]
\begin{center}
  \includegraphics[scale=0.5]{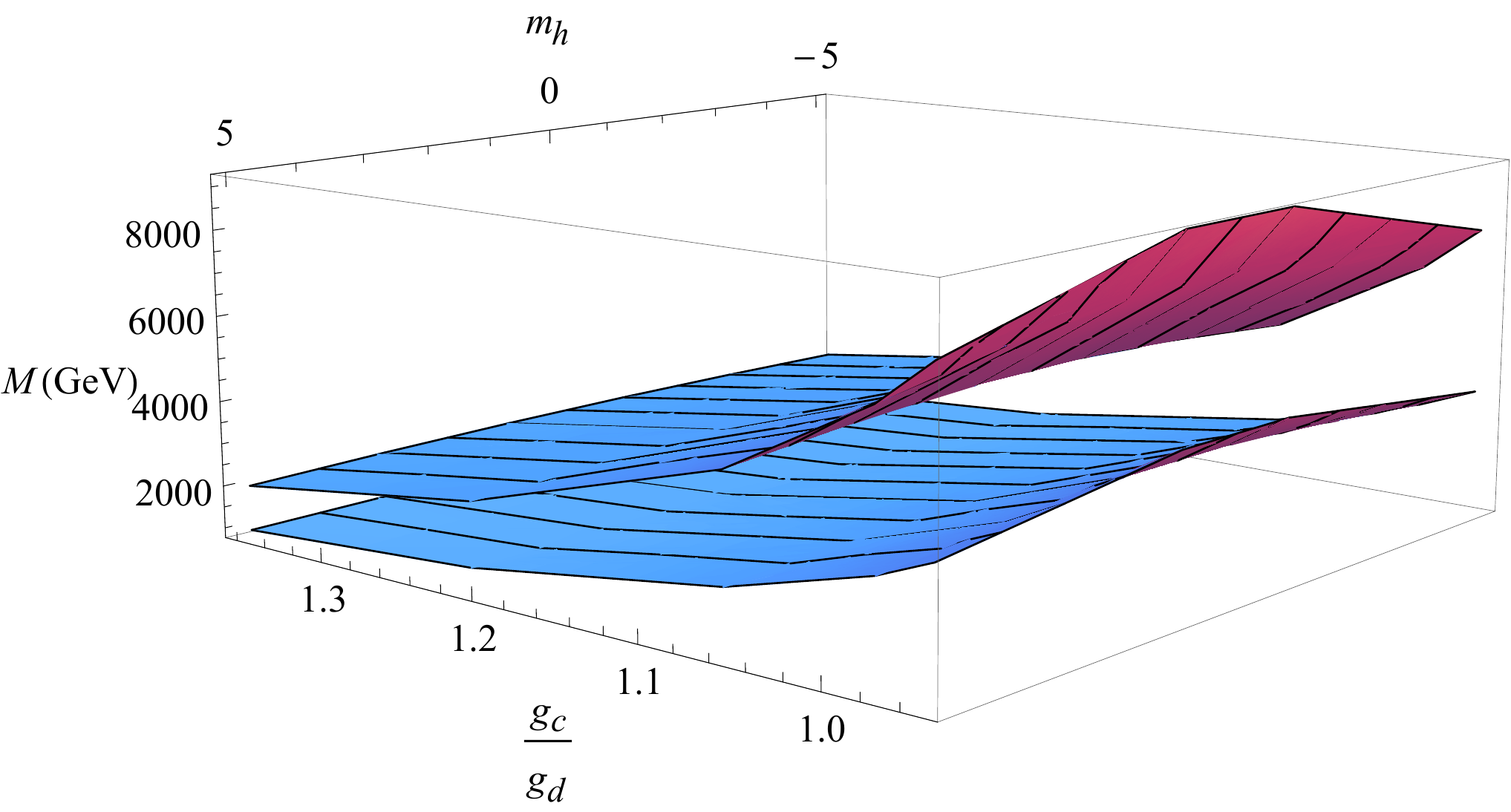}
  \caption{An exhibition of the mass of the $Z'$ (upper surface) and the corresponding dark particle mass (lower surface)
  in terms of $g_{c}/g_{d}$ and $m_{h}$. Again, $m_h$ can only be integer values,
  which are presented as lines on the surfaces.}
  \label{3DDM}
\end{center}
\end{figure}

Given these expressions, one can estimate the expected DM relic density, and compare it with  the Planck result $\Omega_{{\rm DM}}h_{0}^{2}=0.1199\pm0.0027$~\cite{Ade:2013ktc}. 
For illustration, we have shown in figure~\ref{3DDM} the value of the DM masses that lead to the observed relic density exclusively through the $Z'$ annihilation channel, for the explicit toroidal models of section~\ref{model}, assuming that the 12 families of hidden matter fields have the same mass. As one can see, in this case, the estimated mass of the hidden particles is close to half of the $Z'$ mass. This is the region where the Breit-Wigner enhancement is in force, i.e. the region around which the averaged cross section~\eqref{averaged} is peaked, leading to a very efficient DM annihilation~\cite{Griest:1990kh,Gondolo:1990dk,Nath:1992ty}.

\subsection{Hidden valleys and SUSY mediation}
Other than DM physics, one of the motivations for considering hidden sector(s) is supersymmetry (SUSY) breaking. Models of dynamical SUSY breaking typically involve one or more strongly coupled hidden sectors. The strong dynamics of the hidden sector(s) triggers SUSY breaking which is then mediated to the visible sector.  Depending on the scenario, various types of messengers have been proposed.

The St\"uckelberg portal provides a concrete realization of $Z'$ mediation \cite{Langacker:2007ac}, but with key differences in several respects.
The cancellation of $U(1)$ anomalies by the GS mechanism allows us to construct models where the mediation is purely through the $U(1)$ bosons without the need of introducing matter exotics. Furthermore, in  our scenarios the extra $U(1)$ symmetries are not broken by a vev of a scalar field $\langle S \rangle$, but rather through non-perturbative effects (c.f. section~\ref{sec:non-perturbative}). These $U(1)$ symmetries can be used to protect certain operators like the $\mu$-term (or the $\mu B$) or a Dirac neutrino mass, and the pattern of their breaking will affect the resulting phenomenology.

For illustrative purpose, we have presented a class of simple toroidal orientifolds to concretely realize the St\"uckelberg portal scenario. Although these simple models are non-supersymmetric, there is no obstruction in constructing a supersymmetric embedding of our scenario
with intersecting D-branes in Calabi-Yau compactifications. One can then employ the St\"uckelberg portal outlined here to mediate SUSY breaking from the hidden sector to the visible sector.

As in $Z'$ mediation \cite{Langacker:2007ac}, the sfermions from the visible sector couple directly to the $Z'$ messenger, while the gaugino masses are generated only at higher loop order, a typical feature of split supersymmetry~\cite{ArkaniHamed:2004fb,Giudice:2004tc,ArkaniHamed:2004yi}.

Our setup also provides a natural implementation of the hidden valley scenario \cite{Strassler:2006im} (for the defining properties of hidden valley, see section A of \cite{Strassler:2006qa}).
If the hidden sector contains (in addition to $U(1)$'s) a confining non-Abelian gauge group, a mass gap can come e.g. from its strong dynamics.
The barrier energy scale that separates the hidden sector from the visible sector is set by the mass of the lightest $Z'$. The existence of a hidden valley can lead to distinct signatures such as displaced vertices and high multiplicities of jets and leptons in the final state \cite{Strassler:2006im,Strassler:2006qa,Han:2007ae}. 
Here, we note that mass mixing of $U(1)$'s naturally results in such models, and furthermore the choice of $U(1)$'s in such scenarios can be significantly broadened.
For example, the phenomenology of a particular simple case with $U(1)_{\rm v}$ taken as a (anomaly free) linear combination of B-L and hypercharge was explored in detail in \cite{Han:2007ae}. 
The St\"uckelberg mechanism provides a way to cancel the apparent field theoretical $U(1)$ anomalies without introducing chiral exotics.\footnote{In contrast, chiral exotics appeared in other approaches to hidden sectors in string theory~\cite{Cvetic:2012kj,Honecker:2012qr}.} Hence, there are many more choices of $U(1)$'s that one can employ as mediators between the hidden valley and the visible sector.
In this sense, the St\"uckelberg portal described here provides a minimal realization of the hidden valley scenario.

\section{A light $Z'$ from large hidden sectors}\label{sec:random}
As we have seen, the phenomenology of the models we study depend crucially on the mass of the $Z'$ bosons that communicate the hidden and the visible sectors, specially on that of the lightest one. In particular, a too massive $Z'$ would not have a significant impact on the low energy phenomena explored by current or near future experiments. On the other hand, the mass of these $Z'$ bosons seems to be tightly related to the string scale $M_s$. It is important to address the conditions under which a $Z'$ at a phenomenologically relevant scale would arise in our setups. In this section we briefly review the known options to achieve this, and propose a new mechanism to reduce the mass of the lightest $Z'$ bosons with respect to the string scale.

From equation~\eqref{lag}, after reabsorbing the coupling constants from the kinetic matrix $f$ into the gauge bosons $A_a\to g_a A_a$, the mass matrix from the St\"uckelberg mechanism reads:
\begin{equation}\label{massorders}
\tilde{M}^2_{ab}=\sum_{ij}g_a g_b \,K_a^i K_b^j \,G_{ij}\sim {\cal O}(g^2 M_s^2)\,,
\end{equation}
where, $g_a^2\sim g_s/\text{Vol}(\Pi_a)$ are gauge coupling constants inversely proportional to the volumes wrapped by the branes (expressed in string units); $K$ is a matrix of integer charges, and $G\sim M_s^2$ is the positive definite metric of the complex structure moduli space. 

The entries of the mass matrix $\tilde{M}$ are proportional to the string scale $M_s$, so a $Z'$ boson with a TeV mass can be obtained in compactifications with a low string scale~\cite{Ghilencea:2002da}. Because of the relation of the four dimensional Planck mass and the string scale, one must consider a large internal space for such cases. One should notice, however, that in the simple toroidal models of section~\ref{model} it is not possible to arbitrarily increase the volume of the full compactification space. By doing so, one would unavoidably increase the volumes of the cycles wrapped by the visible branes which should be kept fixed since they control the gauge couplings of the SM. The problem comes from the fact that in the simple case of the six-torus, there are no directions in the internal space which are simultaneously perpendicular to all the branes from the visible sector. In order to overcome this caveat one could imagine that the torus is only a small subspace of a large internal manifold, to which it is connected through a throat~\cite{Aldazabal:2000dg,Aldazabal:2000cn,Uranga:2002pg}.

Another way to lower the $Z'$ masses is to consider extraweak couplings $g_a\ll 1$, or equivalently large three cycles $[\Pi_a]$, as can be seen straightforwardly from equation~\eqref{massorders}. Again, the volumes of the visible branes are determined by the values of the SM gauge coupling constants. One could consider anisotropic compactifications in which some of the cycles are small (among them the ones wrapped by visible branes), while some other are large. Wrapping hidden branes along the latter would result in some light $Z'$ bosons in the system~\cite{Goodsell:2009xc}. Despite being an interesting and fruitful method to reduce the mass of some $Z'$ bosons, this mechanism has the unwanted feature of resulting in extraweak couplings to the light $Z'$ bosons (this can be an advantage for some scenarios such as those with `hidden photons' that we briefly discuss in section~\ref{hiddenphotons}).

Here we would like to propose a third possibility, based on the presence of multiple $U(1)$ gauge bosons in the hidden sector, to generate a hierarchy between the $Z'$ and the string mass scales. We have extensively exploited throughout this work the non-diagonal character of the squared mass matrix~\eqref{massorders} to induce mixing among the $U(1)$ bosons. Although each of the entries of $\tilde{M}^2$ is of order $\sim g^2 M_s^2$, if the matrix is large, its lowest eigenvalues may be significantly smaller than that. This well known eigenvalue repulsion effect has been exploited in the context of vacuum stability in the (non-superymmetric) string landscape~\cite{Marsh:2011aa,Chen:2011ac} (see also~\cite{Denef:2004cf}). In this section we would like to perform a preliminary exploration of this effect as a mechanism to lower the mass of the lightest $Z'$ bosons in D-brane models with a large number of $U(1)$'s.

Obviously, if one wants to consider a large number of massive bosons, one must make sure that there are also enough RR axions $\phi^i$ to give a mass to them through the St\"uckelberg mechanism. Since the number of such axions in type IIA string theory is given by the hodge number $h_{2,1}+1$ of the internal CY space, we need to consider for this proposal a compactification in spaces with large $h_{2,1}$. In particular the simple toroidal models we have discussed more explicitly have $h_{2,1}=3$ so they are not suitable for this mechanism.

We study here whether the eigenvalue repulsion effect for the mass matrix can lead to a suppression of the mass of the lightest $Z'$ boson with respect to the string scale $M_s$. We take a statistical random matrix approach to the problem (see e.g. \cite{Stephanov:2005ff} for a review)\footnote{Similar statistical studies of intersecting brane models have also been considered in~\cite{Blumenhagen:2004xx,Gmeiner:2005vz,Douglas:2006xy}.} in which we consider a number $n$ of $U(1)_a$ bosons (i.e. a number $n$ of brane stacks) coupling to $n$ axions $\phi^i$ with random integer charges $K^i_a$ homogeneously distributed within the range $[-10,10]$ (i.e. with random wrapping numbers along the odd cycles $[\beta_i]$). We also take random values of the coupling constants $g_a$ in the range $[-0.001,1]$ also distributed homogeneously. Given the poor control of generic compact CY geometries, there is not much we can say about the complex structure moduli space metric, and so we take for simplicity $G=M_s^2 \times 1_{n\times n}$. In this way, we construct an ensemble of random mass matrices $\tilde{M}^2$.

\begin{figure}[h!t]
\begin{center}
  \includegraphics[scale=1]{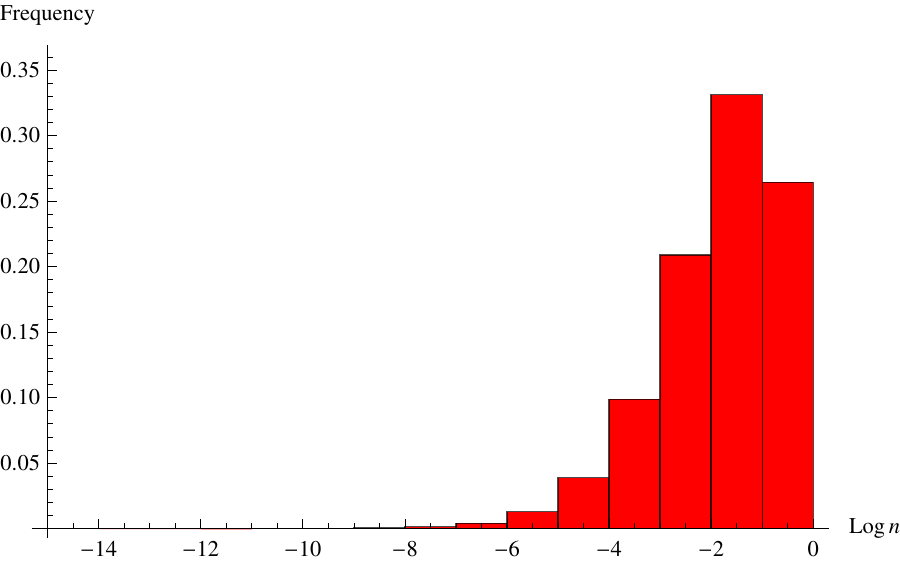}
  \caption{A histogram showing the distribution of the smallest eigenvalue of one million random mass-squared matrices for $n=10$.}
  \label{histogram4}
\end{center}
\end{figure}

Since we are interested here in the properties of massive gauge bosons, we only consider linearly independent charge vectors, i.e. matrices $K$ of rank $n$, so that all the eigenvalues of $\tilde{M}^2$ are positive. As a first step we have considered a sample of one million $10\times 10$ random mass matrices. In the following table and in figure~\ref{histogram4} we show the distribution of their lightest eigenvalues in terms of the string mass $M_s^2$. 

{ \renewcommand{\arraystretch}{1.5}
\begin{center}
\begin{tabular}{|c|c|c|c|c|c|}
\hline
$M_{Z'}^2/M_s^2$ & $10^{-1}-10^{0}$ & $10^{-2}-10^{-1}$ & $10^{-3}-10^{-2}$ & $10^{-4}-10^{-3}$ & $10^{-5}-10^{-4}$\\
\hline
${\rm Frequency}$ & $ 26.42\% $ & $ 33.11\% $ & $20.88\%$ & $9.87\%$ & $3.88\%$\\
\hline\noalign{\smallskip}\noalign{\smallskip}
\hline
$M_{Z'}^2/M_s^2$ & $10^{-6}-10^{-5}$ & $10^{-7}-10^{-6}$ & $10^{-8}-10^{-7}$ & $10^{-9}-10^{-8}$ & $10^{-10}-10^{-9}$\\
\hline
${\rm Frequency}$ & $1.29\%$ & $0.41\%$ & $0.13\%$ & $0.04\%$ & $0.01\%$\\
\hline\noalign{\smallskip}\noalign{\smallskip}
\hline
$M_{Z'}^2/M_s^2$ & $10^{-11}-10^{-10}$ & $10^{-12}-10^{-11}$ & $10^{-13}-10^{-12}$ & $10^{-14}-10^{-13}$ & $0-10^{-14}$\\
\hline
${\rm Frequency}$ & $47\times 10^{-6}$ & $11\times 10^{-6}$ & $4\times 10^{-6}$ & $2\times 10^{-6}$ & $0$\\
\hline
\end{tabular}
\end{center}}

Unfortunately, we see from these results that the eigenvalue repulsion effect is not too large. The majority of events yield a  ratio for $M_{Z'}^2/M_s^2$ in the range $(10^{-6},1)$, which can be easily explained by the range of gauge couplings $(0.001,1)$ chosen. Nevertheless, events below the $10^{-6}$ threshold can only be attributed to the eigenvalue repulsion, and we see that there is a non-zero chance to obtain a $Z'$ mass as low as $10^{-6}M_s$.

\begin{figure}[h!t]
\begin{center}
  \includegraphics[scale=0.8]{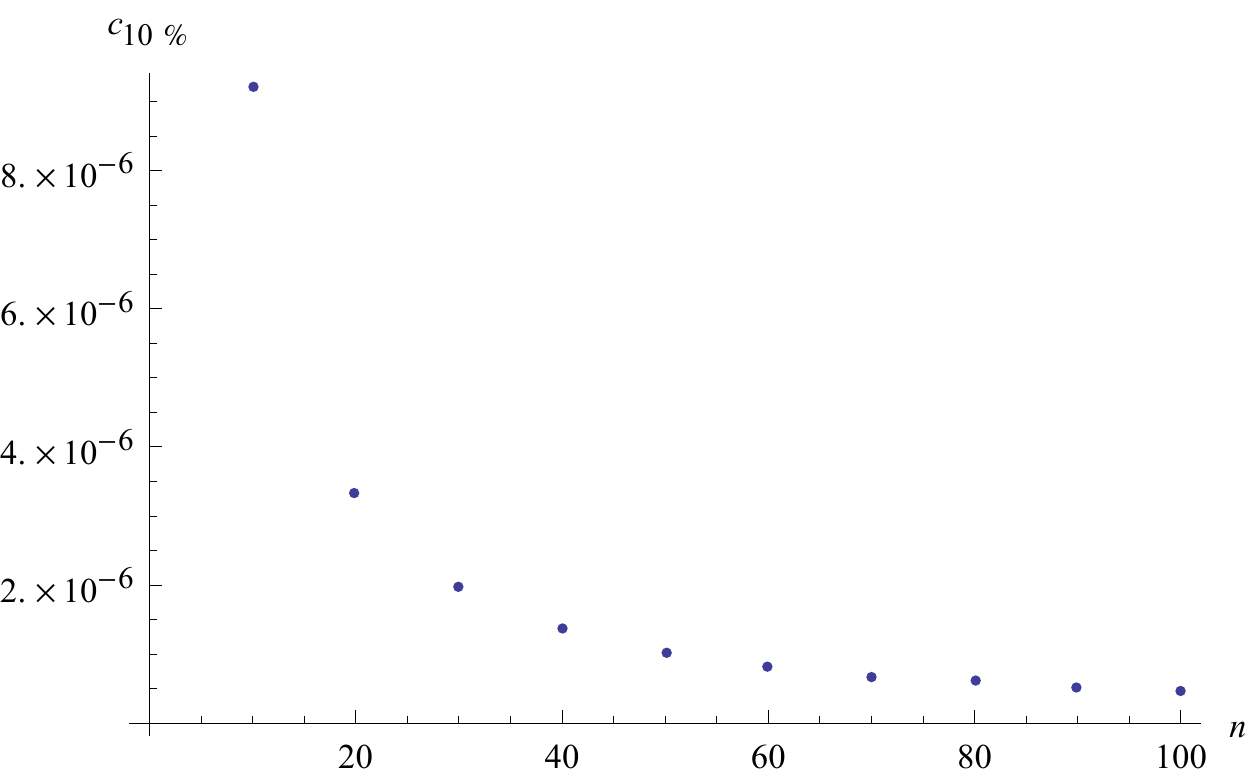}
  \caption{An exhibition of $c_{10\%}$ as a function of $n$. The best fitting function reads
  $c_{10\%} = 2.33\times 10^{-6} \, n^{-1.41}$.}
  \label{RM10p}
\end{center}
\end{figure}

It is also interesting to study how these results depend on the size $n$ of the mass matrix, i.e. on the number of $Z'$ bosons in the system. To this end, we randomize matrices $K$ of different sizes $n=10,20,\ldots,100$, as well as the gauge coupling constants $g_a$ in the range $[-0.001,1]$, and compute the lightest eigenvalue of the mass-squared matrix $\tilde{M}^2$.
The results of a run with one million events are summarized in the following table (where  $M^2_{Z'} = c M^2_s$) and in figure~\ref{RM10p}.
{ \renewcommand{\arraystretch}{1.5}\begin{center}
\begin{tabular}{|c|c|c|c|c|c|}
\hline
$n$ & $10$ & $20$ & $30$ & $40$ & $50$\\
\hline
$c_{50\%}$ & $3.7\times10^{-4}$ & $1.2\times10^{-4}$ & $6.5\times10^{-5}$ & $4.3\times10^{-5}$ & $3.2\times10^{-5}$\\
\hline
$c_{10\%}$ & $9.2\times10^{-6}$ & $3.3\times10^{-6}$ & $2.0\times10^{-6}$ & $1.3\times10^{-6}$ & $1.0\times10^{-6}$\\
\hline
$c_{1\%}$ & $9.1\times10^{-8}$ & $3.3\times10^{-8}$ & $1.9\times10^{-8}$ & $1.2\times10^{-8}$ & $1.0\times10^{-8}$\\
\hline\noalign{\smallskip}\noalign{\smallskip}\noalign{\smallskip}\noalign{\smallskip}
\hline
$n$ & $60$ & $70$ & $80$ & $90$ & $100$\\
\hline
$c_{50\%}$ & $2.5\times10^{-5}$ & $2.1\times10^{-5}$ & $1.8\times10^{-5}$ & $1.6\times10^{-5}$ & $1.3\times10^{-5}$\\
\hline
$c_{10\%}$ & $8.1\times10^{-7}$ & $6.8\times10^{-7}$ & $6.0\times10^{-7}$ & $5.0\times10^{-7}$ & $4.5\times10^{-7}$\\
\hline
$c_{1\%}$ & $7.1\times10^{-9}$ & $7.0\times10^{-9}$ & $5.5\times10^{-9}$ & $5.1\times10^{-9}$ & $4.7\times10^{-9}$\\
\hline
\end{tabular}
\end{center}}
Here $c_{50\%}$ is the median number of the set, which means that
$50\%$ of the events in this set have a smaller value than $c_{50\%}$.
The numbers $c_{10\%}$ and  $c_{1\%}$ are defined in a similar way.

We can see that the masses of the lightest $Z'$ do indeed decrease as the number of $U(1)$ bosons increase. This behavior is due exclusively to the eigenvalue repulsion. Although the dependence on $n$ and the mass suppression are relatively mild, we believe that the statistical large $n$ effect is an interesting mechanism which can be added to the suppressions by relatively small gauge coupling constants and fundamental string scale to explore scenarios with $Z'$ bosons in a mass range relevant for current phenomenological studies.

Our rough discussion serves to illustrate the effect that a large number of branes has on the mass of the lightest $Z'$ bosons, but it is evident that it should be supplemented by further analysis in several aspects. In particular, one would like to implement these models in explicit geometries and study the effect of the complex structure metric $G$ in the resulting mass eigenvalues. Also, we have not implemented in this section the tadpole cancellation conditions~\eqref{tadpoles} which must be satisfied in any complete model.

Finally, in a realistic compactification one would need to include a visible sector with branes that lead to the SM gauge group and field content, e.g. by realising the intersection numbers~\eqref{intersm}. Such a sector could be connected through the St\"uckelberg portal to a hidden sector with numerous brane stacks where the large $n$ effect we have discussed takes place. One would have to make sure that there is no exotic chiral matter charged under the SM, i.e. that the visible branes do not intersect any hidden ones. Since we are describing setups with a large hodge number $h_{2,1}$, these conditions can generically be implemented as in the simple example presented in section~\ref{model}. The conditions can be easily considered case by case, although a systematic implementation in the random matrix approach requires more involved algorithms that we leave for future work. 

In any case, the effects of the tadpole cancellation condition as well as the inclusion of a visible sector are expected to be statistically irrelevant in the large $n$ limit.

\section{Related portals}\label{relatedportals}

In this last section we would like to comment on two other mechanisms, both related to the St\"uckelberg portal, which also connect the visible and the hidden sectors.

\subsection{Non-perturbative effects}\label{sec:non-perturbative}
In the type of constructions we have discussed in this work, there is another type of interaction between the hidden and the visible sectors, besides those mediated by gravity and the studied $Z'$ bosons. In fact, it is also induced by the RR-axions $\phi^i$ involved in the St\"uckelberg portal we have discussed, but it manifests itself as a non-perturbative coupling that effectively breaks the massive $U(1)$ symmetries (for a review on non-perturbative effects in type II theories see \cite{Blumenhagen:2009qh}). Similar effects have been discussed in the context of SUSY mediation in~\cite{Buican:2008qe}.

Recall from eq.\eqref{axionshift} that the operators $e^{-U^i}=e^{-u^i-i\phi^i}$ transform linearly under $U(1)_a$ gauge transformations, with a vector of charges given by $(K)_a^i=N_a r_a^i$. As we have discussed, the St\"uckelberg portal arises when, for some $\phi^i$, this vector contains non-zero charges for both visible and hidden $U(1)$ symmetries. 

In this case, one can construct $U(1)$-invariant operators that mix visible and hidden charged fields, i.e. an operator of the form $e^{-U^i}\Psi_{\text{v}}\Psi_{\text{h}}$, where $\Psi_{\text v}$ and $\Psi_{\text h}$ are operators from the visible and hidden sectors whose $U(1)$ charges cancel those of $e^{-U^i}$.

Indeed, such type of operators are generated by instantonic D2-branes that wrap around the cycle $[\alpha^i]$ and have a classical action given by $U^i$, and whose non-perturbative contribution to the effective action is weighted precisely by $e^{-U^i}$. The fact that these instantons generate interactions between charged operators is related to the appearance of charged fermionic zero-modes on the worldvolume of the instanton, that arise from the intersections of $[\alpha^i]$ with the gauge branes of the system. As is well known, upon gauge fixing and stabilization of the complex structure moduli, such effects are responsible for the non-perturbative breaking of the $U(1)$ symmetries that become massive through the St\"uckelberg mechanism. \cite{Blumenhagen:2006xt,Ibanez:2006da,Florea:2006si}

The particular operators generated this way depend on all the zero-modes on the worldvolume of the instanton brane, and must be analyzed case by case for different compactifications. Since the effects of such interactions are non-perturbatively suppressed and expected to be small in comparison with those mediated by $Z'$ bosons, we will not study them further. Nevertheless, it is worth noting that both type of interactions between separated sectors, St\"uckelberg and non-perturbative, are tightly connected.

\subsection{`Hidden photon' scenarios}\label{hiddenphotons}

So far, we have considered in this work scenarios where a $Z'$ boson couples with significant strengths both to the visible and the visible sectors. As we have seen, this setup can be achieved by non-diagonal terms in the mass matrix $M^2=(K^{\,T}\cdot G\cdot K)$ induced by mixed axionic charges encoded in the matrix of integers $K$. 

It is also interesting to consider a different possibility, in which light gauge bosons have a significant coupling to the hidden sector while they have extremely weak interactions with the visible sector. Because of this small coupling, the severe restrictions on the mass of the $Z'$ bosons from direct searches (c.f. section~\ref{sec:exp}) do not apply, and one can consider so called `hidden photon' scenarios, where the mass of such bosons can be extremely light.

The usual way to introduce such a setup is to consider a hidden $U(1)$ boson $A_h$ with a mass $m_h$ that has a small  kinetic mixing $\delta\ll1$ with hypercharge $A_Y$ (or equivalently with the photon):
\begin{equation}
{\cal L}=-\frac{1}{4g^2_Y}F_Y^2-\frac{1}{4g^2_h}F_h^2-\frac{\delta}{2}F_YF_h-\frac{1}{2}m_h^2 A_h^2 +A_Y J_Y+A_h J_h
\end{equation}
One can get rid of the mixing and set a diagonal kinetic matrix (while keeping the mass matrix diagonal) by redefining $A_Y\to A_Y-\delta g_y^2 A_h$. In this way, the hidden boson $A_h$, that originally coupled to the hidden current $J_h$ acquires a small coupling to the visible hypercharge coupling $J_Y$. Visible matter fields interact with the hidden photon with a strength proportional to $\delta$ and their hypercharge. Such scenarios arise naturally in string constructions~\cite{Lust:2003ky,Abel:2003ue,Berg:2004ek,Abel:2006qt,Abel:2008ai,Goodsell:2009xc}.  
Their phenomenology has been thoroughly studied for various ranges of the parameters $\delta$ and $m_h$ (see~\cite{Jaeckel:2010ni,Essig:2013lka} and references therein). 

In this section, we want to point out that a similar scenario can be obtained through small non-diagonal terms in the mass (rather than kinetic) matrix in the presence of multiple $U(1)$ factors. While mixings induced by the integer matrix $K$ are generically large, a small mixing can be induced by small non-diagonal elements in the matrix $G$, i.e. by small kinetic mixings between the axions absorbed by the $U(1)$ bosons.

In order to avoid large mixings that would induce large couplings of gauge bosons to separated sectors, we assume that there are no axions charged simultaneously under both. In the language of section~\ref{sec:theory}, for every axion $\phi^i$ the vector of $U(1)$ charges $k_a^i$ has non-zero entries for $a$ either in the visible sector or the hidden sector, but not for both simultaneously (that is, $K$ takes a block diagonal form). 

A small mixing between axions that couple to different sectors can be induced by non diagonal terms in the moduli space metric $G$ (i.e. the kinetic matrix of the axions), see figure~\ref{mixingdiagram}. We can construct a toy model, with two massive gauge bosons (different from hypercharge), one in the visible sector $A_v$ and one in the hidden one $A_h$. They couple via a St\"uckelberg term to two axions $\phi^v$ and $\phi^h$ respectively with a matrix of charges $K=1_{2\times2}$. The mixing is induced by a non-diagonal small term in the axionic kinetic matrix that we take of the form
\begin{equation}\label{modmetric}
G=\left(\begin{array}{cc}
m_v^2&\epsilon\\
\epsilon &m_h^2\end{array}\right)\,,
\end{equation}
with $\epsilon\ll m_h^2<m_v^2$. The effective Lagrangian involving the gauge bosons would read
\begin{equation}\label{hiddenphotonlag}
{\cal L}=-\frac{1}{4g^2_v}F_v^2-\frac{1}{4g^2_h}F_h^2-\frac{1}{2}m_v^2 A_v^2-\frac{1}{2}m_h^2 A_h^2 -\epsilon A_vA_h +A_v J_v +A_h J_h\,,
\end{equation}
where we have neglected kinetic mixing effects.\footnote{Since we are considering small effects in this section, kinetic mixings would be important and should be included in a complete discussion. For our purpose of illustrating the mass mixing effect it is sufficient to consider the simplified Lagrangian~\eqref{hiddenphotonlag}.} Again, we can get rid of the mixing by a field redefinition which, to first order in $\epsilon$ reads
\begin{equation}
A_v\to  A_v - \tilde{\epsilon}A_h\,,\qquad \qquad A_h\to  A_h + \tilde{\epsilon}A_v\,,
\end{equation}
where we have defined 
\begin{equation}
\tilde{\epsilon}\equiv \frac{g_vg_h\epsilon}{g_v^2m_v^2-g_h^2m_h^2}\,.
\end{equation}
After this redefinition, the Lagrangian in terms of the physical bosons reads
\begin{equation}
{\cal L}\approx-\frac{1}{4g^2_v}F_v^2-\frac{1}{4g^2_h}F_h^2-\frac{1}{2}m_v^2 A_v^2-\frac{1}{2}m_h^2 A_h^2 +A_v( J_v+\tilde{\epsilon}J_h) +A_h (J_h-\tilde{\epsilon}J_v)\,.
\end{equation}
From this, we see that the hidden current $J_h$ couples to the visible gauge boson $A_v$, and at the same time the visible current $J_v$ couples to the hidden photon $A_h$, both with a small coupling controlled by $\tilde{\epsilon}$. 

Since the visible boson $A_v$ couples to the visible sector with significant strength, its mass must satisfy the bounds described in section~\ref{sec:exp}, i.e. we should consider $m_h$ at the multi TeV scale. On the contrary, $A_h$ couples very weakly to the visible sector, and is a perfect candidate for a hidden photon, whose mass may be extremely small. Such small masses could be achieved in principle by applying to the hidden sector some of the mechanisms described in section~\ref{sec:random}.

The phenomenology of this scenario is very similar to that of the models considered in the literature in which a hidden photon mixes kinetically with hypercharge. An important difference is that, in our case, the coupling of matter fields to the hidden photon is not proportional to its hypercharge, but depends on its charges under $U(1)_v$. Since we are dealing with a St\"uckelberg  $U(1)$, this symmetry may be anomalous, with anomalies cancelled through the generalized GS mechanism. This leads to a richer class of hidden photon scenarios.

It would be interesting to implement this class of scenarios in an explicit string compactification. In principle, a moduli space metric such as~\eqref{modmetric} could arise from quantum corrections to a metric which is diagonal at tree level, such as that of a torus (c.f. equation~\eqref{Gtorus}). Unfortunately, corrections to the complex structure moduli space are very hard to compute and at the present stage it is not possible to realize our proposal in generic compactifications.

\section{Conclusions}\label{conclusions}

In this paper, we have presented a natural framework for the visible sector to interact with the
hidden sector(s). We have shown how  Stueckelberg $U(1)$'s can provide an interesting portal into  hidden sector(s). Our scenario can be viewed purely field-theoretically, though it 
is particularly well motivated from string theory. We have explicitly constructed a class of intersecting brane models to illustrate how this Stueckelberg portal scenario can be embedded into string theory. These intersecting brane models are {\it global} constructions (i.e., tadpole free) which extend the Madrid quiver (where the Standard Model reside) with a {\it genuine} hidden sector, i.e., there are no chiral exotics with charges under the Standard Model. We have carried out some preliminary  phenomenological studies of this scenario, pointing out some model-independent features as well as presenting results for some concrete D-brane models to illustrate our approach. The phenomenological features of interest include
  $Z-Z'$ mixings, DM stability and relic density, hidden valleys and SUSY mediation, etc. We have also discussed related portals, emphasizing the differences from previous works. We pointed out
  that in addition to kinetic mixing assumed in previous literature, mass mixing
  can also  have significant effects in the
  `hidden photon' scenario. 
  This leads to a broader class of `hidden photon' models, where the coupling of the hidden photon with the visible sector is distinct from what was previously studied in this context.
  
The Stueckelberg portal proposed here bears some similarities with 
other $Z'$ mediation scenarios. However there are some notable differences. In particular, the Stueckelberg portal has several added appealing features that are advantageous from both phenomenological and model building standpoints. 
Among these new features are (i) a broader choice of $U(1)$'s can be made without the need of introducing unwanted exotics, and (ii) a more sizable interaction between the visible and hidden sector can arise as the mixings are generated at tree-level rather than loop suppressed.

As a phenomenological scenario, the mass of the lightest Stueckelberg $Z'$ is a free parameter. However, in string constructions, the $Z'$ mass matrix depends on the string scale $M_s$ (the only fundamental energy scale in the problem), the associated gauge couplings, and the axion charges under the $U(1)$'s. We pointed out that in addition to lowering the string scale or considering ultra-weak hidden sector couplings, eigenvalue repulsion in the mass matrix provides  yet another way to lower the mass of the $Z'$s. Given an ensemble of large rank matrices, one can ask what is the likelihood of finding a significantly reduced $Z'$ mass.
Our preliminary study 
indicated that increasing the number of branes (i.e., $U(1)$'s) can help lower the mass of the lightest $Z'$ but clearly further analysis is needed.
For example, it would be interesting to 
implement the data for 
explicit geometries and study the effect of the complex structure metric $G$ in the resulting
mass eigenvalues. It is also important to see if the inclusion of
a visible sector with the SM gauge group and field content affect the statistical statements we made here.
A
systematic implementation of these additional features and the corresponding constraints (e.g., tadpole cancellation, absence of exotics) in this statistical approach requires more involved algorithms
that we leave for future work.

\noindent{\bf Acknowledgments:}
We are grateful to  H.~An, P.~C\'amara, N.~Chen, I.~Garc\'ia-Etxebarr\'ia, M.~Goodsell, J.~Hajer, L.~Iba\~nez, D.~Junghans, T.-J.~Li, T.~Liu, Z.~Liu, L.-C.~L\"u, F.~Marchesano, P.~Nath, R.~Richter, Y.~Sumitomo, H.~Tye, A.~Uranga, W.~Xue and H.~Zhang for helpful discussions. G.S. and P.S. would like to thank the Instituto de F\'isica Te\'orica of Madrid for kind hospitality. 
This work 
is supported in part by the DOE grant DE-FG-02-95ER40896 and the HKRGC grant 604213. WZF is also supported by the Alexander von Humboldt Foundation.

\printbibliography

\end{document}